\documentclass{article}
\usepackage{xcolor}
\usepackage{hyperref}
\usepackage{graphicx}
\usepackage{blindtext}
\usepackage{fullpage}
\usepackage{comment}
\usepackage{amsmath}

\usepackage[adobe-utopia]{mathdesign}
\usepackage[T1]{fontenc}

\graphicspath{ {./imgs} }

\newcommand{\rev}[1]{#1}

\title{The geometry of efficient codes: how rate-distortion trade-offs distort the latent representations of generative models}

\author{Leo D'Amato  \\
	Polytechnic University of Turin  \& \\
 Institute of Cognitive Sciences and Technologies, National Research Council, Rome, Italy \\
	\and 
	Gian Luca Lancia \\
	Institute of Cognitive Sciences and Technologies, National Research Council, Rome, Italy  \&  \\
    Department of Psychology, Sapienza University of Rome, Rome, Italy \\
    \and
Giovanni Pezzulo\footnote{Corresponding author: giovanni.pezzulo@istc.cnr.it}$^*$  \\
	Institute of Cognitive Sciences and Technologies, National Research Council, Rome, Italy  \\ 
	}

\begin{document}

\maketitle

%\textcolor{red}{uniformare i termini usati: latent codes, latent representations, latent states, embeddings}

\begin{abstract}
    Living organisms rely on internal models of the world to act adaptively. These models, \rev{because of resource limitations,} cannot encode every detail and hence need to compress information. From a cognitive standpoint, information compression can manifest as a distortion of latent representations, resulting in the emergence of representations that may not accurately reflect the external world or its geometry. Rate-distortion theory formalizes the optimal way to compress information \rev{while minimizing such distortions}, by considering factors such as capacity limitations, the frequency and the utility of stimuli. However, while this theory explains why the above factors distort latent representations, it does not specify which specific distortions they produce. To address this question, here we \rev{investigate how rate-distortion trade-offs shape the latent representations of images in generative models, specifically Beta Variational Autoencoders ($\beta$-VAEs), under varying constraints of model capacity, data distributions, and task objectives. By systematically exploring these factors, we identify three primary distortions in latent representations: prototypization, specialization, and orthogonalization. These distortions emerge as signatures of information compression, reflecting the model's adaptation to capacity limitations, data imbalances, and task demands. Additionally, our findings demonstrate that} these distortions can coexist, giving rise to a rich landscape of latent spaces, whose geometry could differ significantly across generative models subject to different constraints. Our findings contribute to explain how the normative constraints of rate-distortion theory \rev{shape} the geometry of latent representations of generative models of artificial systems and living organisms.
    
\end{abstract}


\section*{Author summary}

To understand and act adaptively in the external world, living organisms rely on compressed internal models due to the limitations of their cognitive resources. In turn, information compression results in distortions of their internal representations, causing them to differ from the actual external world. While distortions of internal models have been reported in both artificial systems and living organisms, they are still poorly understood. In this study, we investigate adaptive solutions to information compression, under the normative principle of rate-distortion theory. We identify three primary distortions — prototypization, specialization, and orthogonalization — that emerge under various constraints on capacity, data distributions, and tasks. Our findings offer a framework for understanding these geometric distortions in latent representations, potentially shedding light on the causes of distortions in the neural coding of living organisms.

\section{Introduction}\label{sec:intro}

Living organisms rely on internal models to navigate their environment effectively \cite{de2022predictive, epstein2017cognitive}. However, due to the high metabolic and processing costs involved, these models cannot capture every detail. Therefore, organisms must compress information adaptively to operate efficiently \cite{laughlin1998metabolic}. The brain is believed to utilize efficient neural representations for perceptual processing, aiming to maximize information content while minimizing metabolic expenditure \cite{barlow1961possible,olshausen1996emergence}. Evidence of adaptive information compression spans various cognitive domains, including perception \cite{bates2020efficient, attneave1954some, barlow1974redundancy}, working memory \cite{jakob2023rate, desimone1995neural, brady2009compression, mathy2012s, yoo2018strategic}, cognitive mapping \cite{lloyd1989cognitive,tversky1992distortions}, and decision making \cite{bhui2021resource,lewis2014computational, gershman2015computational, griffiths2015rational}.

In efficient coding, information compression relies on at least three key factors. The first and most apparent factor is the system's capacity limitation, often measured by the number of bits it can encode. The other two factors pertain to the frequency and utility of stimuli. Stimuli that occur more frequently and are instrumental in achieving goals are prioritized for accurate encoding and thus allocated more resources. Conversely, stimuli of lower frequency or lesser importance can be encoded with fewer resources and less fidelity \cite{tishby2000information,burge2015optimal}.

From a cognitive standpoint, information compression can manifest as a distortion of latent representations — resulting in the emergence of representations that may not accurately reflect the external world or its geometry. Multiple independent lines of research have investigated how capacity limitations, stimulus likelihood, and their utility influence cognitive representations. It has been long recognized that capacity limitations can lead to distortions in latent representations, such as spatial models or cognitive maps \cite{lloyd1989cognitive,tversky1992distortions}. For instance, in geographic maps, there is a tendency towards shape compactness with a regression towards the mean of extreme borders, producing a prototype-like effect \cite{costa2018geometrical}. In addition to spatial mapping, during decision-making, humans often exhibit systematic distortions in representations of value and probability. These distortions are thought to be approximately optimal given the assumption that decisions are made with finite computational resources \cite{juechems2021optimal}. Other distortions in cognitive processing are contingent upon the likelihood of stimuli. For instance, working memory often exhibits systematic biases towards stimuli that are more frequently presented than others \cite{panichello2019error}. Certain aspects of neural coding, such as burst coding and firing rate adaptation, can be seen as characteristic features of optimal coding strategies aimed at facilitating accurate inference \cite{mlynarski2018adaptive}.

Moreover, distortions can also hinge on the utility of stimuli for task-solving or goal achievement. For example, during goal-directed navigation, rewards have been observed to compress spatial representations of mazes in both the hippocampus and orbitofrontal cortex \cite{muhle2023goal}. In the early stages of sensory processing, neural representations of stimuli optimize not only for information but also for fitness, particularly when different stimuli are associated with distinct rewards \cite{schaffner2023sensory}. Some distortions may dynamically emerge depending on the current goal or context. For instance, the prefrontal cortex is known to remap representations of perceptually identical items based on their utility in achieving specific goals \cite{castegnetti7usefulness}. Additionally, attention mechanisms tend to enhance neural representations of changes in sensory input at the expense of perceptual accuracy \cite{mehrpour2020attention}. Finally, learning a novel task has been observed to render visual cortical population codes more orthogonal \cite{failor2021visuomotor}. Taken together, these and various other studies underscore the adaptive nature of information compression in the brain.

\begin{figure}
\begin{center}
\includegraphics[width=\columnwidth]{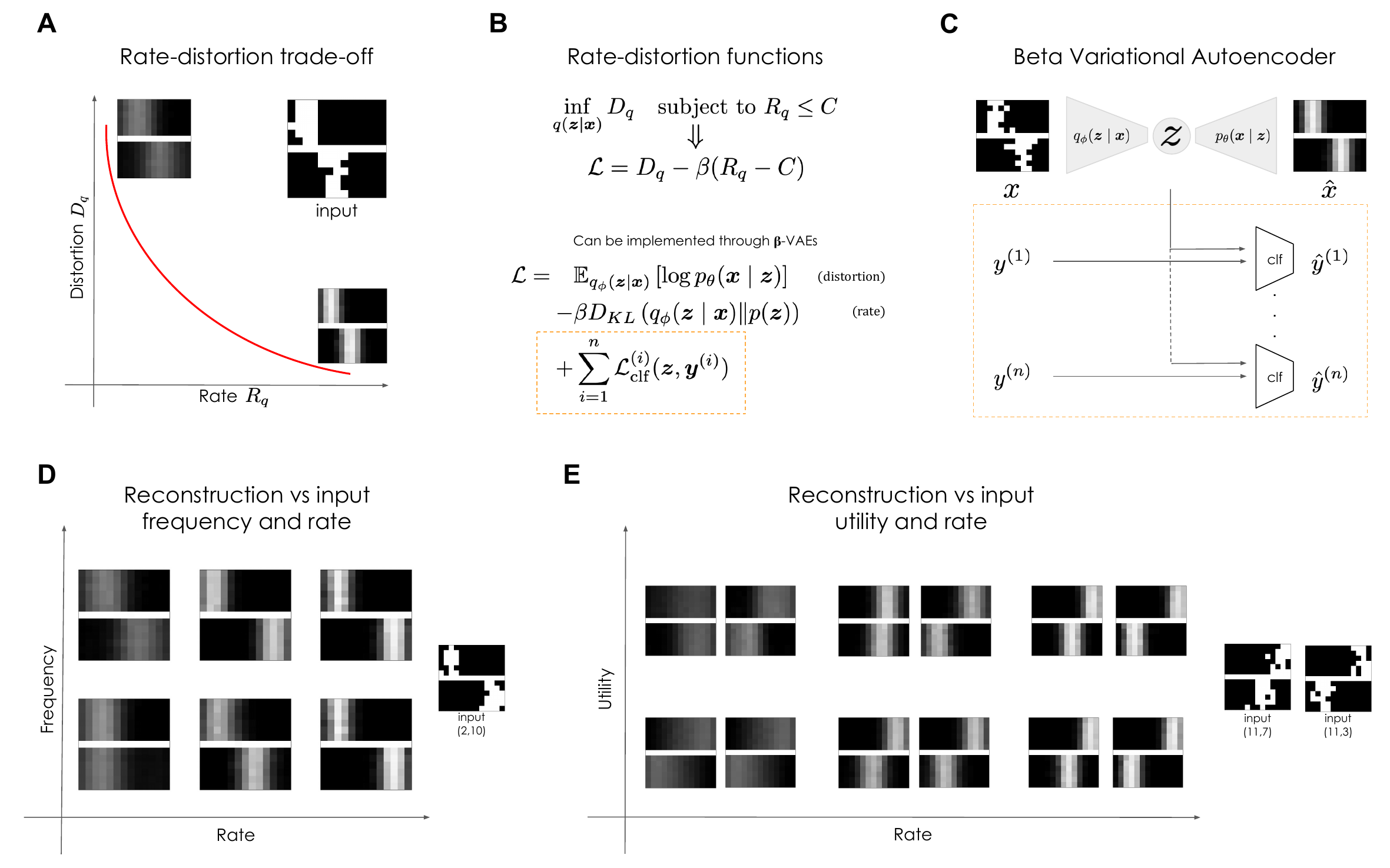}
\caption{Formal framework adopted in this study, based on rate-distortion theory (RDT) \cite{shannon1959coding}. (A) Trade-off between \emph{rate} (or available resources) and \emph{distortion}. (B) The goal of RDT is to find the minimum distortion function given a constraint on the available resources. The loss of a $\beta$ variational autoencoder has the same shape as the Lagrangian of this minimization problem. (C) Architecture of the network used in experiments: it is a classical $\beta$-VAE that can be optionally augmented with $n$ classifiers. In most of our experiments $n$ is equal $0$ or $1$. The classifiers can be both linear or non-linear. The loss of the classifiers is added to the loss of the $\beta$-VAE as described in panel B. (D) Stimuli with higher utility are encoded more faithfully. Utility of a stimulus is related to its likelihood or to its relevance with respect to a task. This panel shows that, under strong resource constraints (small rate), stimuli with small probability of occurrence are ignored. (E) This panel shows how faithfully a stimulus is reconstructed under different rate and relevance conditions: at small rates, two stimuli with small relevance are collapsed into the same representation while details about stimuli with high relevance are still preserved even at small rates.}
\label{fig:intro1}
\end{center}
\end{figure}

These findings can be framed within the context of rate-distortion theory \cite{shannon1959coding}, providing a normative framework for understanding adaptive information compression in communication channels and potentially in the brains of living organisms. Rate-distortion theory elucidates why factors such as information capacity, data likelihood, and utility for goal attainment can distort latent representations by formalizing the trade-off between information rate (the average number of bits per stimulus used for encoding) and distortion (the cost of reconstruction errors). Figure \ref{fig:intro1}A illustrates the rate-distortion trade-off, depicting the relationship between information rate (the average number of bits per stimulus transmitted across the memory channel) and distortion (the cost of memory errors). Systems with higher capacity tend to achieve lower expected distortion, delineating an optimal trade-off curve in the rate-distortion plane. For instance, when reconstructing the same input image, distortion is more pronounced (i.e., the reconstructed image appears blurry or averages out features from the dataset) under low resource availability, while it is minimized (i.e., the reconstructed image is highly accurate) under conditions of ample resources. Several studies have demonstrated the utility of rate-distortion theory in explaining various cognitive phenomena, including working memory  \cite{jakob2023rate,sims2012ideal, sims2015, nagy2020optimal}, perceptual processing \cite{sims2016rate,sims2018efficient,sims2003implications}, category learning \cite{bates2019adaptive}, visual search \cite{bates2021optimal}, and decision making \cite{gershman2020origin,lai2021policy}.

However, while rate-distortion theory provides a normative principle to understand why latent representations are distorted, we still lack a systematic understanding of how they are distorted under different training conditions—specifically, with varying capacity, data distributions, tasks/goals, or their combinations, and how these distortions interact. 

\begin{figure}[t!]
\centering
\includegraphics[width=\columnwidth]{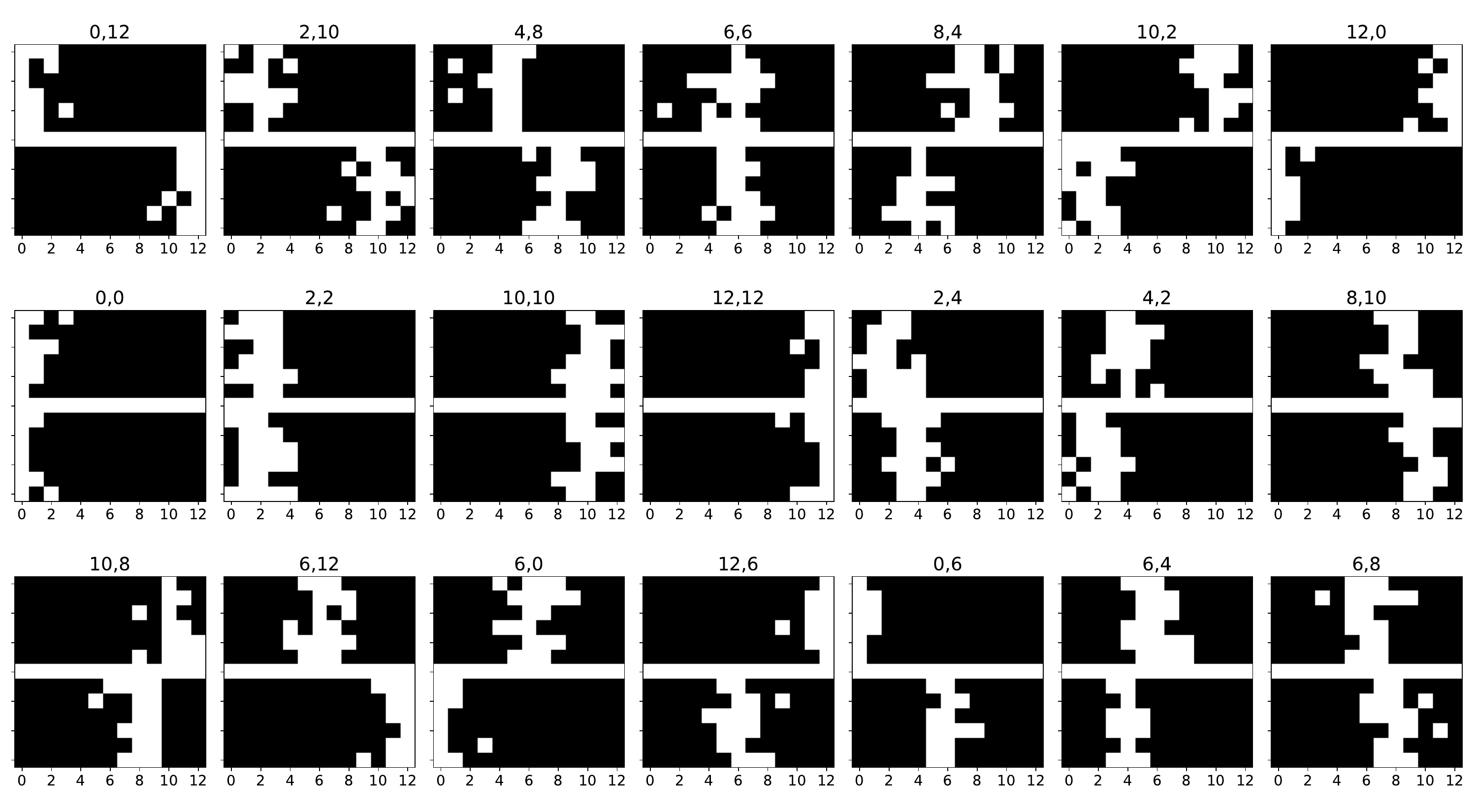}
\caption{The \emph{Corridors} dataset. The figure shows 21 example images in the ``corridors'' dataset used for this study. Each figure comprises two white corridors, placed at the upper and at the lower parts of the image, with a white horizontal line that is common to all the images. Each corridor is a noisy vertical line of white pixels, whose true center is in one of 13 $x$ positions, coded from 0 (left) to 12 (right). For example, in the first image, the upper corridor is in the position $x_{UC}=0$ and the lower corridor is in the position $x_{LC}=12.$. The corridor positions are reported on top of each image.}
\label{fig:dataset}
\end{figure}

To address this challenge, here we systematically examine the \rev{distortions in the} geometry of latent representations (embeddings) \rev{of generative models, specifically Beta Variational Autoencoders ($\beta$-VAEs), under varying constraints of model capacity, data distributions, task objectives or their combinations.} We focus on two sets of experiments that involve memorizing sets of images \rev{representing} simple spatial maps composed of two ``corridors'', placed at the upper and lower parts of the image, respectively (Figure \ref{fig:dataset}). The first experiment allows us to test the distortions of latent representations that arise under different capacities and data distributions, such as familiar or unfamiliar images. In this experiment, models are trained to remember images from balanced or unbalanced datasets, where the to-be-remembered image appears more or less frequently. The second experiment enables us to test the distortions of latent representations emerging under different capacities and tasks/goals. Here, models are trained to perform various image classification tasks, such as determining whether the upper corridor is to the left or right of the bottom corridor, or whether the corridors are aligned or not. For instance, for the first image in Figure \ref{fig:dataset}, the correct answers would be ``yes'' in the first case and ``no'' in the second case.

To conduct the experiments, we employ a Beta Variational Autoencoder ($\beta$-VAE) \cite{bvae_2017}, \rev{since it} has been demonstrated to approximate rate-distortion theory \cite{park2022interpreting, bvae_2018}. Moreover, it allows for the utilization of real-world stimuli such as images \cite{bates2020efficient,nagy2020optimal}. The $\beta$-VAE learns in an unsupervised manner to compress information and create latent internal representations or \emph{embeddings} during the encoding phase (memory encoding or perceptual learning) and its reconstruction accuracy is evaluated during the decoding phase (memory retrieval or generation). In the second experiment set, the $\beta$-VAE is supplemented by a series of classifiers, each trained in a supervised manner to address a specific classification task.

\rev{The dataset, the code and the models' checkpoints of our experiments are available at: \\ \href{https://github.com/damat-le/geom_eff_codes}{https://github.com/damat-le/geom\_eff\_codes}.}

\section{Results}

\subsection{Problem specification: the two experiments}

Here, we perform two experiments. \emph{Experiment 1} addresses the distortions induced by statistically biased training sets by comparing the representations learned by the same neural network, first trained on an unbiased dataset, and then trained on a biased one. \emph{Experiment 2} addresses the distortions induced by the classification task assigned to the network. To do this, we compared the representation learned by the network when the only objective was to reconstruct input images against the representation learned when the objective was augmented with a classification task. In both experimental sets, we trained multiple networks by varying their encoding capacity to study how this impacts on the magnitude of distortions (see below).
\\

\subsection{The Corridors dataset}

To study the emergence of disentangled or distorted latent representations in generative models, we designed a novel dataset -- the Corridors dataset -- in which two \rev{generative} factors vary orthogonally. 

The Corridors dataset consists of $200.000$ black and white 13x13 images\rev{. Each image is divided into two sections, upper and lower, by an horizontal white line of width 1 pixel. Each section includes a vertical noisy ``corridor'', i.e., a white vertical core segment with white random pixels around it (Figure \ref{fig:dataset}). The white random pixels are independent samples of a Bernoulli distribution $f(g(x;\mu, \sigma))$, where $g(x;\mu, \sigma)$ is a Gaussian probability density function with standard deviation $\sigma=1.2$ and mean $\mu$ equal to the position of the vertical core segment. The position of the vertical core segment, also referred to as the position of the corridor, is an integer number ranging from 0 to 12, since each corridor can slide horizontally from pixel 0 to pixel 12 across images. Thus, each image is associated with two labels, $(x_{UC}, x_{LC})$, where $x_{UC}$ denotes the position of the upper corridor and $x_{LC}$ denotes the position of the lower corridor. The positions of the upper and lower corridors represent the two orthogonal factors of variation of the dataset. The Corridors dataset is available at: \href{https://github.com/damat-le/geom_eff_codes}{https://github.com/damat-le/geom\_eff\_codes}.
}

%containing 2 vertical noisy ``corridors'' (i.e., columns of white pixels), one in the upper part of the image and the other one in lower part (Figure \ref{fig:dataset}). In each image, the upper and lower sections are separated by an horizontal \rev{white} line with width 1 pixel. The positions of the upper corridor ($x_{UC}$) and of the lower corridor ($x_{LC}$) are two integers that represent the two orthogonal factors of variation of the dataset, since they can slide horizontally from pixel 0 to pixel 12 across images. Each pixel of the image was obtained as an independent sample from a Bernoulli distribution $f(g(x;\mu, \sigma))$, where $g(x;\mu, \sigma)$ is a Gaussian probability density function with standard deviation $\sigma=1.2$ and mean $\mu=x_{UC}$ if the pixel is in the upper half of the image, $\mu=x_{LC}$ otherwise.

\subsection{Experiment 1: distortions induced by varying model capacity and data distributions}\label{sec:exp1} 

In Experiment 1, we examine the distortions of the $\beta$-VAE representations, as an effect of model capacity and statistically biased training sets, in which some types of images are more frequent than others. 

For this, we train 3 models: baseline, E1M1, and E1M2. The baseline $\beta$-VAE model is trained on the Corridors dataset. The E1M1 model is a $\beta$-VAE trained on an unbalanced dataset, in which the images with the lower corridor on the left appear 10 times more often than images with the lower corridor on the right. Let $x_{LC}$ be the pixel at which the lower corridor is centered, then ``lower corridor on the left'' means that $x_{LC} \leq 6$, where $x_{LC} \in \{0,1,\ldots,12\}$. The E1M2 model is a $\beta$-VAE trained on an unbalanced dataset in which the images having the two corridors aligned ($x_{LC} = x_{UC}$) appear 10 times more often than images with non-aligned corridors.

We repeat the training of all the models (baseline, E1M1, E1M2) for 5 different values of encoding capacity $C_{\operatorname{max}} \in \{0.3, 1, 3, 6, 10\}$ nats. 

\subsubsection{Performance of the models}

\begin{figure}
\begin{center}
\includegraphics[width=\columnwidth]{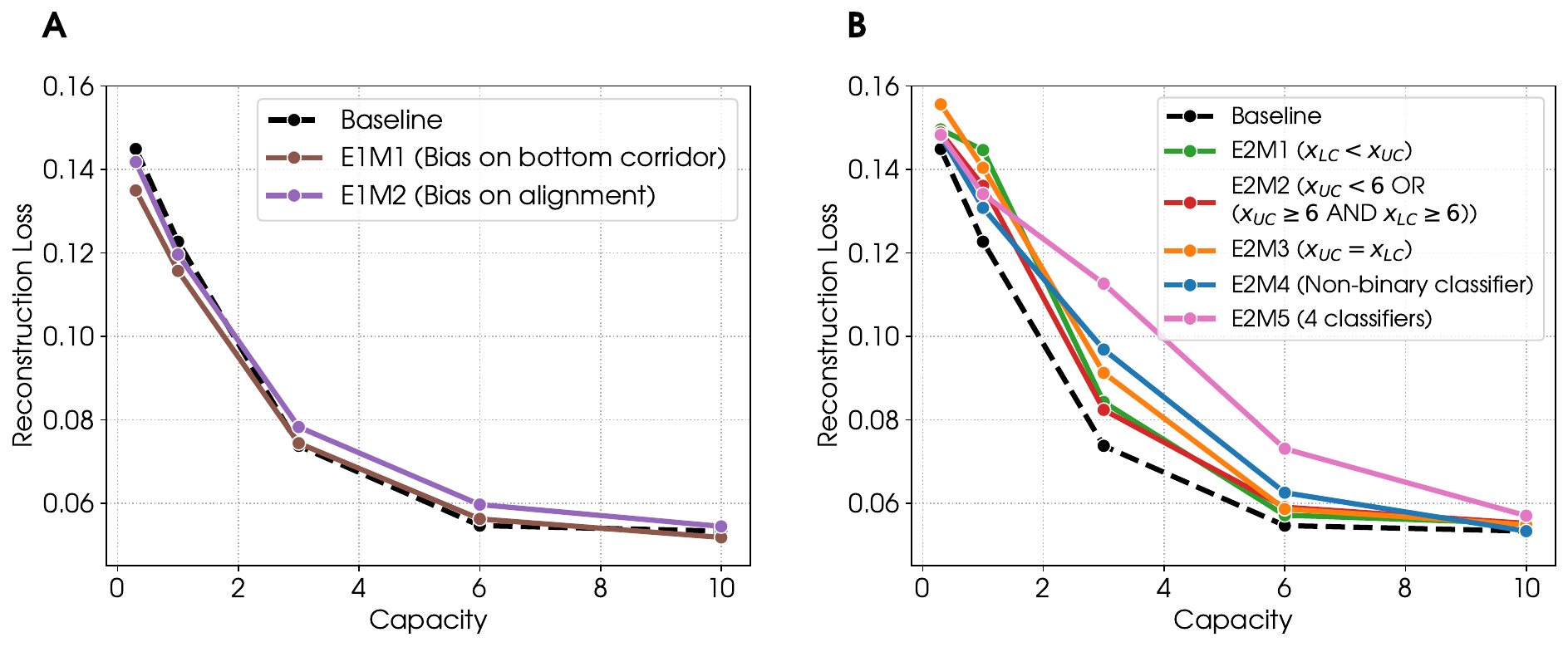}
\caption{Reconstruction loss, all the models. \rev{For ease of reading, a brief description of each model along with its label is reported in the legend.} (A) Experiment 1. The figure illustrates that increasing capacity reduces reconstruction loss. The trend is similar for the baseline $\beta$-VAE model that is trained with a balanced dataset and for the two models (E1M1 - E1M2) that are trained with unbalanced datasets. (B) Experiment 2. The figure illustrates that at any given capacity, the baseline model has a smaller reconstruction loss compared to the hybrid models that are additionally trained to solve classification tasks (E2M1 - E2M5). Furthermore, at any given capacity, reconstruction loss changes across the different tasks and is worst for the E2M5 model, which addresses four classification tasks simultaneously. See the main text for explanation.}
\label{fig:rec_loss_all}
\end{center}
\end{figure}

Figure~\ref{fig:rec_loss_all}A shows the reconstruction loss of the three $\beta$-VAE models trained on the balanced dataset (baseline), the unbalanced dataset with a bias for lower corridor to the left (E1M1), and the unbalanced dataset with a bias for corridors aligned (E1M2), at the 5 different capacities of $C_{\operatorname{max}} \in \{0.3, 1, 3, 6, 10\}$ nats. As expected, for all models, reconstruction loss decreases as the model's capacity increases. The decrease of reconstruction loss follows the same trend across all models.

\subsubsection{Geometry of the latent representations of the models}

We next compare the representations learned by the three models, when trained at high ($C_{\operatorname{max}} = 10$ nats) and low capacity ($C_{\operatorname{max}} = 0.3$ nats). 

\paragraph{Baseline model, at high and low capacity}

\begin{figure}
\begin{center}
\includegraphics[width=\columnwidth]{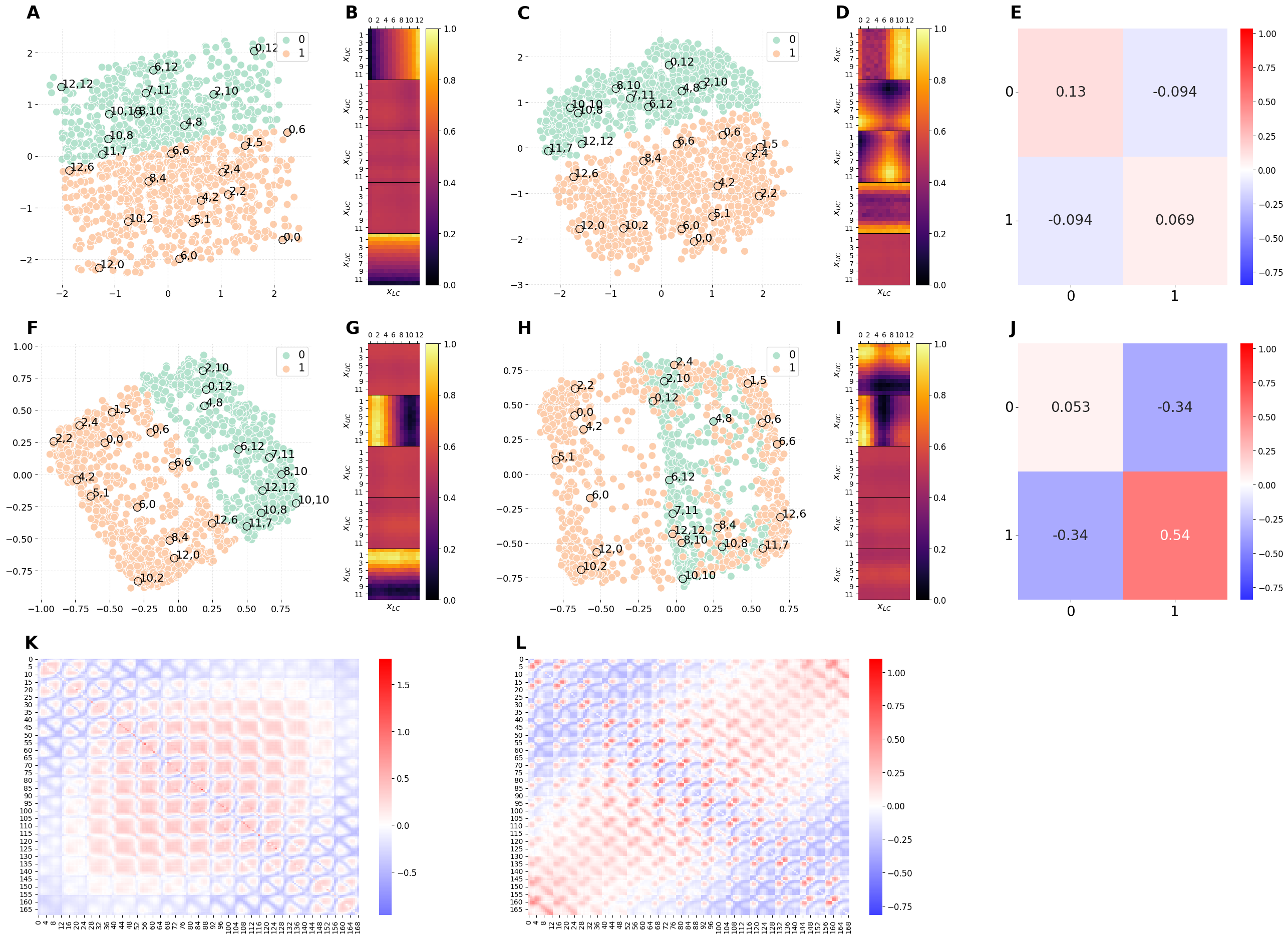}
\caption{
Comparison of the latent representations of the baseline model and the model E1M1. The E1M1 model is trained on an unbalanced dataset, in which images with $x_{LC} \leq 6$ (orange) are 10 times more frequent than images with $x_{LC} > 6$ (green).
\textbf{(A, F)} 2D projections of the 5D embeddings learned by the baseline model at high ($C_{\max}=10$ nats) and low ($C_{\max}=0.3$ nats) capacity, respectively.
\textbf{(C, H)} 2D projections of the 5D embeddings learned by the hybrid E1M1 model trained at high ($C_{\max}=10$ nats) and low ($C_{\max}=0.3$ nats) capacity, respectively. 
\textbf{(B, G)} Activation patterns of the 5 latent channels of the baseline model at high and low capacity, respectively. Each of the five heat-maps is computed as in sec. \ref{sec:neuron_activity}. 
\textbf{(D, I)} Activation patterns of the 5 latent channels of the E1M1 model at high and low capacity, respectively. Each heat-map is computed as in Section~\ref{sec:neuron_activity}. 
\textbf{(E, J)} Measure of distortions in the latent representations of the hybrid E1M1 model compared to the baseline model at high and low capacity, respectively. 
\textbf{(K)} Measure of distortions in the latent representation of the baseline model induced by the reduction of the encoding capacity from high to low.
\textbf{(L)} Measure of distortions in the latent representation of the E1M1 model induced by the reduction of the encoding capacity from high to low.
Distortion matrices are computed as described in Section~\ref{sec:compare_lr}.
}
\label{fig:E1M1}
\end{center}
\end{figure}

Figures~\ref{fig:E1M1}A and F show 2D projections of the latent spaces learned by the baseline $\beta$-VAE model trained on the balanced dataset, at high and low capacity, respectively. In these and the subsequent figures, each dot corresponds to the position of a specific image within the low-dimensional latent space of the generative model. For ease of interpretation, some images are labeled with corridors positions, such as "6,6" in the center. The colors of the dots correspond to the classes of images that have a high frequency (green) or low frequency (orange) in the E1M1 and E1M2 models. These colors are irrelevant for the baseline model, but are retained for ease of comparison with the E1M1 and E1M2 models, see later.

Figure~\ref{fig:E1M1}A-B shows that, at high capacity, the representation is perfectly \emph{disentangled}. It is possible to notice that the geometry of the latent space is a square, with the two axes corresponding to the two generative factors of variation of the dataset: namely, the positions of the two corridors. Moving along one axis while keeping the other fixed affects the position of only one of the two corridors. For example, moving from the top-left to the top-right angles of the square corresponds to moving from the image (0,0) to the image (12,0). In other words, the first factor (the position of the top corridor, from 0 to 12) changes, whereas the second factor (the position of the bottom corridor, from 0 to 12) remains the same. The opposite is true if one moves from the top-left to the bottom-left angles of the square. The disentanglement of the internal code is confirmed by the activation patterns of the latent channels, shown in Figure \ref{fig:E1M1}B. The figure shows that only two channels are active, each of which encoding the position of a single corridor (the color changes only along one of the two axes). Moreover, the color transition is smooth, indicating that the model is able to encode every single value of the position of each corridor.

\begin{figure}
\begin{center}
\includegraphics[width=\columnwidth]{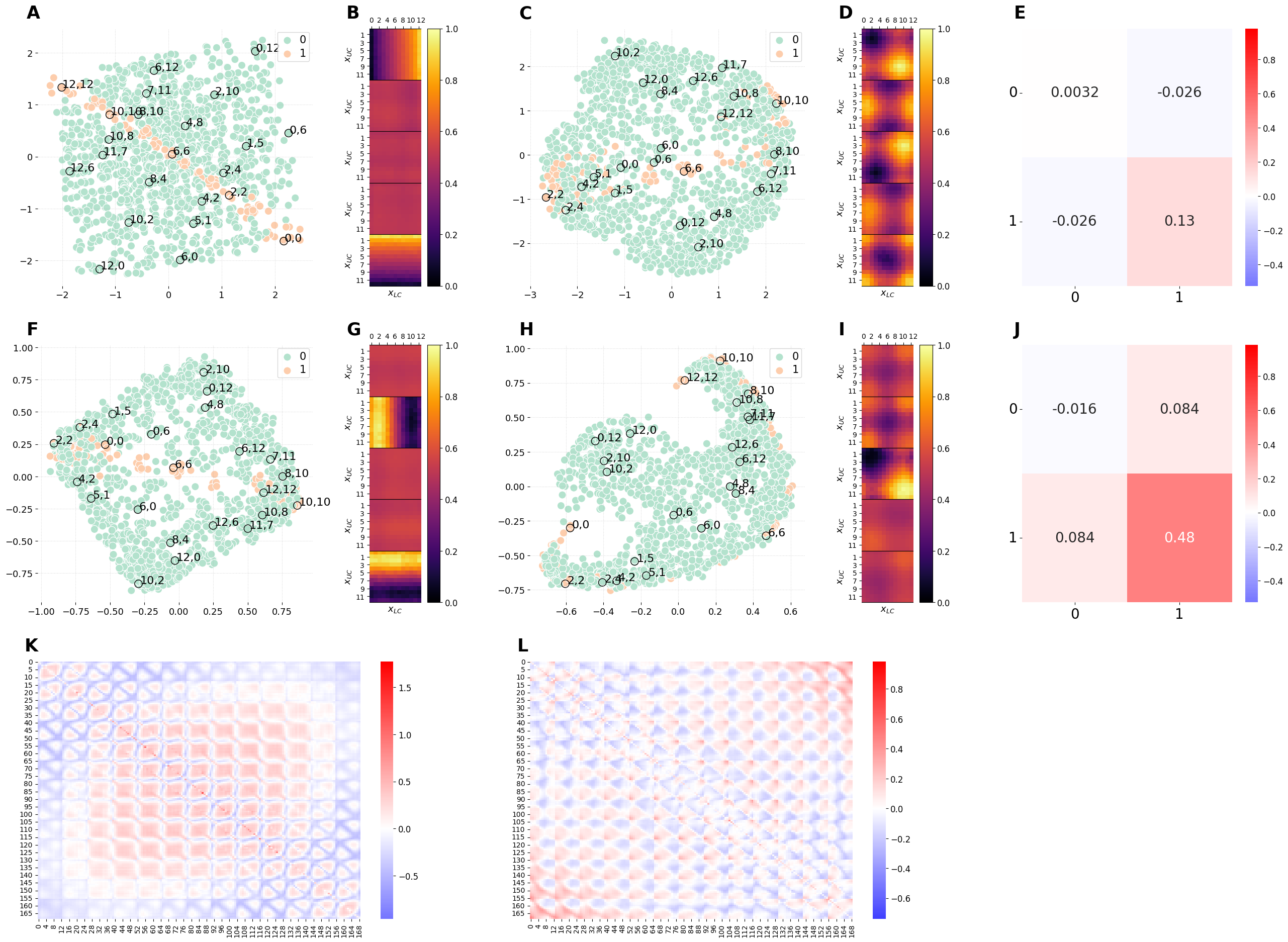}
\caption{
Comparison of the latent representations of the baseline model and the hybrid model E1M2. The E1M2 model is trained on an unbalanced dataset in which images with $x_L = x_U$ (orange) are 10 times more frequent than images with $x_L \neq x_U$ (green). % The network is assigned a binary classification task, in which images with $x_U \leq x_L$ are labeled as $1$, $0$ otherwise.
\textbf{(A, F)} 2D projections of the 5D embeddings learned by the baseline model at high ($C_{\max}=10$ nats) and low ($C_{\max}=0.3$ nats) capacity, respectively.
\textbf{(C, H)} 2D projections of the 5D embeddings learned by the hybrid E1M2 model trained at high ($C_{\max}=10$ nats) and low ($C_{\max}=0.3$ nats) capacity, respectively. 
\textbf{(B, G)} Activation patterns of the 5 latent channels of the baseline model at high and low capacity, respectively. Each of the five heat-maps is computed as in sec. \ref{sec:neuron_activity}. 
\textbf{(D, I)} Activation patterns of the 5 latent channels of the E1M2 model at high and low capacity, respectively. Each heat-map is computed as in Section~\ref{sec:neuron_activity}. 
\textbf{(E, J)} Measure of distortions in the latent representations of the hybrid E1M2 model compared to the baseline model at high and low capacity, respectively.
\textbf{(K)} Measure of distortions in the latent representation of the baseline model induced by the reduction of the encoding capacity from high to low.
\textbf{(L)} Measure of distortions in the latent representation of the E1M2 model induced by the reduction of the encoding capacity from high to low.
Distortion matrices are computed as described in Section~\ref{sec:compare_lr}.
}
\label{fig:E1M2}
\end{center}
\end{figure}

Figure~\ref{fig:E1M1}F-G shows that when baseline model is trained at low capacity, the representation is still disentangled, as evident from the fact that it uses two latent channels for two orthogonal dimensions of the task (Figure \ref{fig:E1M1}G). Furthermore, the latent space maintains a square-like shape (Figure~\ref{fig:E1M1}F), even if the scales of the two axes is much smaller than the in Figure~\ref{fig:E1M1}A and some points are flipped between the two latent spaces (e.g., 0,12 and 10,10). Importantly, we observe a \emph{prototypization} effect: the model collapses the representations of images that are similar to one another. This is evident from the fact that the data points are clustered around the vertices of the square that correspond to four main type of images (or prototypes): both corridors on the left (2,2), both corridors on the right (10,10), upper corridor on the left and lower corridor to the right (2,10), upper corridor on the right and lower corridor on the left (10,2). The prototypization effect is also apparent when comparing the activation patterns of the latent channels in the baseline model at high (Figure \ref{fig:E1M1}B) and low (Figure \ref{fig:E1M1}G) capacity. At low capacity, the color transitions are no longer smooth as for the model at high capacity; rather, each channel presents a more binarised activation as if it is able to encode only two possible positions of the corresponding corridor. The differences between the high and low capacity baseline models are quantitatively measured in Figure \ref{fig:E1M1}K in terms of dilation and compression. Considering the ordering of images as described in Section \ref{sec:compare_lr}, we observe that, on average, pairs of images at the center of the square of Panel A are pushed farther apart when reducing the capacity. Note that the observations made in this paragraph about panels A, B, F, G and K hold true for the Figures \ref{fig:E1M1}, \ref{fig:E1M2}, \ref{fig:E2M1}, \ref{fig:E2M2}, \ref{fig:E2M3}, \ref{fig:E2M4} and \ref{fig:E2M5}. In fact, they all represent the same baseline model, except for the fact that that data points have been assigned different colors to facilitate comparison with other models.

\paragraph{Comparison of baseline and E1M1 models}

Figures \ref{fig:E1M1}C and \ref{fig:E1M1}H show 2D projections of the latent spaces learned by the $\beta$-VAE trained on the unbalanced dataset with a bias for the bottom corridor to the left (E1M1), at high and low capacity, respectively. At high capacity (Figure \ref{fig:E1M1}C), the data imbalance slightly alters the structure of the latent space and it is still possible to recognise a square-like shape. The high similarity between the latent spaces learned at high capacity with balanced and unbalanced datasets is quantified in Figure~\ref{fig:E1M1}E. Despite this similarity, the activation patterns of the latent channels for the unbalanced dataset (Figure \ref{fig:E1M1}D) differ substantially from the baseline model, with four active channels. The first and the fourth channels encode the position of a single corridor in a binarised fashion (this explains the square-like shape and the two clusters). Rather, the second and third channels enrich the latent representation with fine-grained details about corridors positions, which allow the network to reconstruct more faithfully the input stimuli.

At low capacity (Figure \ref{fig:E1M1}H), the number of active channels returns back to two (one per corridor), as shown in Figure \ref{fig:E1M1}I. In particular, the first channel encodes upper corridor positions while the  the second channel only encodes bottom corridor positions $\leq 6$, (i.e. the most frequent images), ignoring images with bottom corridor position >6 (i.e., the least frequent images); note that the color corresponding to $x_{LC} >6$ is equal to the color of inactive channels. This still results in a square-like shape of the latent space, but in this case the square only comprises the most frequent images. The less frequent images are ignored in the latent representation and when the network is presented with a image of type $(x_{UC}, x_{LC})$ with $x_{LC} > 6$, it generates an image of type $(x_{UC}, 3)$. This explains why the less frequent images are positioned at the center of the square. In conclusion, at low capacity, the network encodes the most frequent images (with relatively high fidelity), but not the less frequent images -- plausibly, because ignoring rare images still ensures a small loss. We call this a \emph{specialization} effect, in the sense that the model is biased in the allocation of resources. The specialization effect is similar to the prototypization, but is specific for frequent stimuli rather than forming averages. The magnitude of such specialization effect with respect to the baseline model at low capacity is reported in Figure \ref{fig:E1M1}J while the magnitude of distortions arising from capacity constraints in the E1M1 model are Figure \ref{fig:E1M1}L. Finally, Supplementary Figure~\ref{fig:E1M1_3D} permits appreciating in more detail the latent representations acquired by the E1M1 model, by showing them in 3D.

\paragraph{Comparison of baseline and E1M2 models}

Figures \ref{fig:E1M2}A and \ref{fig:E1M2}F show 2D projections of the latent spaces learned by the $\beta$-VAE trained on the balanced dataset (baseline), at high and low capacity, respectively. These are identical to Figures \ref{fig:E1M1}A and \ref{fig:E1M1}F discussed above, except that the colors of the dots indicate the images that have high frequency (aligned corridors in orange) or low frequency (non-aligned corridors in green) in the E1M2 model. 

Figures \ref{fig:E1M2}C and \ref{fig:E1M2}H show 2D projections of the latent spaces learned by the $\beta$-VAE trained on the unbalanced dataset with a bias for aligned corridors (E1M2), at high and low capacity, respectively. When the E1M2 model is trained with high capacity, both the latent space (Figure~\ref{fig:E1M2}C) and the latent channels (Figure~\ref{fig:E1M2}D) differ significantly from the baseline model (Figure~\ref{fig:E1M2}A). However, Figure~\ref{fig:E1M2}E shows that despite this significant reconfiguration of latent space geometry, the mutual distances between points do not change significantly, except for points belonging to class 1 (aligned corridors). This indicates that the E1M2 model manages to faithfully represent the most frequent images (aligned corridors), by building an ad-hoc latent representation for them -- and then it allocates the remaining capacity to encode the other, less frequent images. As a result, images with aligned corridors are more dilated in the latent space of the E1M2 model with respect to the baseline model. The consequence of such distortion is that all the other images (class 0) change position in the latent space, but still preserve their mutual distances as in the baseline model. Notably, this reconfiguration requires using more than two latent channels (Figure~\ref{fig:E1M2}D), which implies that the latent space is higher dimensional compared to the baseline model.

Similar considerations hold for the E1M2 trained with low capacity (Figure~\ref{fig:E1M2}H). However, in this case the \emph{specialization} effect -- and the significant compression required to encode the points belonging to class 1 (aligned corridors) with sufficient fidelity -- moves them farther apart in latent space compared to the baseline model (Figure~\ref{fig:E1M2}J). As a consequence, the overall latent space appears significantly reconfigured and folded (Figure~\ref{fig:E1M2}H). The folding can be appreciated in greater detail by considering Supplementary Figure~\ref{fig:E1M2_3D}, which shows the latent representation in 3 dimensions. Finally, Figure \ref{fig:E1M2}L shows that reducing model capacity induces a complex pattern of distortion in the latent representation of the E1M2 model.  

\subsubsection{Summary of the results of Experiment 1}

Experiment 1 explored the geometry of latent space that emerge in $\beta$-VAE models, by varying model capacity (i.e., high or low capacity) and data distributions (i.e., balanced vs. unbalanced datasets).

To summarize, our results show that that a baseline $\beta$-VAE trained with a balanced data distribution correctly recovers the two generative factors of variation of the dataset (i.e., the positions of the upper and lower corridors, which vary independently). This is evident from the fact that the geometry of the latent states is a square and the two factors of variation are encoded separately, in two latent channels. When the  baseline model is trained at low capacity, we found that the model performance decreases (as expected) and the disentanglement coexists with a prototypization effect: namely, the model collapses the representations of images that are similar to one another. 

Rather, the models E1M1 and E1M2 trained on unbalanced datasets develop a latent representation that is distorted compared to the baseline model and this distortion is more evident at low capacity. In this case, the models E1M1 and E1M2 show \emph{specialization}: they devote more resources to encode the most common stimuli and this changes the position of their corresponding latent compared to the baseline model. Interestingly, not only training on unbalanced datasets disrupts the disentanglement but it also renders the latent representations higher-dimensional. This is plausibly because it is easier to distinguish common from rare stimuli in more dimensions.

\subsection{Experiment 2: distortions induced by varying model capacity and tasks}\label{sec:exp2}

In Experiment 2, we examine the distortions of the $\beta$-VAE embeddings, as an effect of varying model capacity and including classification tasks. 

For this, we compare the same $\beta$-VAE baseline model considered above with five models: E2M1, E2M2, E2M3, E2M4, and E2M5. The first four models (E2M1 - E2M4) are hybrid models composed of a $\beta$-VAE and one classifier (Fig \ref{fig:intro1}B). These hybrid models are trained in a supervised manner on the same dataset. However, for each model the images are relabeled, depending on the classification task that the model has to solve. The classification tasks assigned to the four hybrid models are explained below:

\begin{enumerate}
    \item The E2M1 model is assigned a binary classification task. We labeled with $1$ the images in which $x_{UC} \leq x_{LC}$ and with $0$ all the other images. In other words, the correct class is $1$ if the upper corridor is to the left of or aligned with the right corridor, and $0$ otherwise.
    \item The E2M2 model is assigned a binary classification task. We labeled with $1$ the images in which $x_{UC} <6$ (regardless of the value of $x_{LC}$) OR $( x_{UC} \geq 6$ AND $x_{LC} \geq 6 )$ and with $0$ all the other images. 
    \item The E2M3 model is assigned a binary classification task. We labeled with $1$ the images in which the corridors are aligned ($x_{LC} = x_{UC}$) and with $0$ all the other images. 
    \item The E2M4 model is assigned a classification task with 25 classes. The 25 classes are defined as follows: we divided both upper and lower part of the image in 5 bins respectively. The upper corridor is assigned a label from 0 to 4 according to which bin it occupies. The bin size is not uniform and the bins for corridor positions are as follows: [0,1] (bin 0), [2,4] (bin 1), [5,7] (bin 2), [8,10] (bin 3), [11, 12] (bin 4). The same holds for the lower corridor. Each image is then assigned one of the 25 possible combinations of labels according to the positions of the two corridors. For example, an image with $(x_{UC}, x_{LC}) = (6, 11)$ has the upper corridor falling into bin 2 and the lower corridor into bin 4 and the it is assigned the label $2*5 + 4 = 14$.
\end{enumerate}
S
Finally, E2M5 is a hybrid model composed of a $\beta$-VAE and 4 classifiers, trained to solve the 4 above tasks simultaneously.

For all the five hybrid models (E2M1 - E2M5), the $\beta$-VAE and classifier(s) are trained jointly end-to-end in a supervised manner on the same dataset, whose images are labeled depending on the task. All the classifiers are linear, except the one used in the E2M4 model, which is non-linear, since a linear classifier is not sufficient to solve the task (see Section \ref{sec:models}). 

We repeat the training of all the models (E2M1, E2SM2, E2M3, E2M4, E2M5) for 5 different values of encoding capacity $C_{\operatorname{max}} \in \{0.3, 1, 3, 6, 10\}$ nats. 

\subsubsection{Performance of the models}

Figure~\ref{fig:rec_loss_all}B shows the reconstruction loss of the baseline $\beta$-VAE model and the hybrid $\beta$-VAE - classifier models, trained in each of the four classification tasks (E2M1 - E2M4) and in all the four classification tasks simultaneously (E2M5). As expected, in all cases, reconstruction loss decreases as the model's capacity increases. Furthermore, as expected, the reconstruction loss is better for the baseline model compared to the hybrid models, since the latter have to optimize an additional (classification) objective. The decrease of reconstruction loss follows the same trend in all models, but the loss remains slightly higher for the model that solves the four classification tasks simultaneously (E2M5).

\begin{figure}[h]
\begin{center}
\includegraphics[width=\columnwidth]{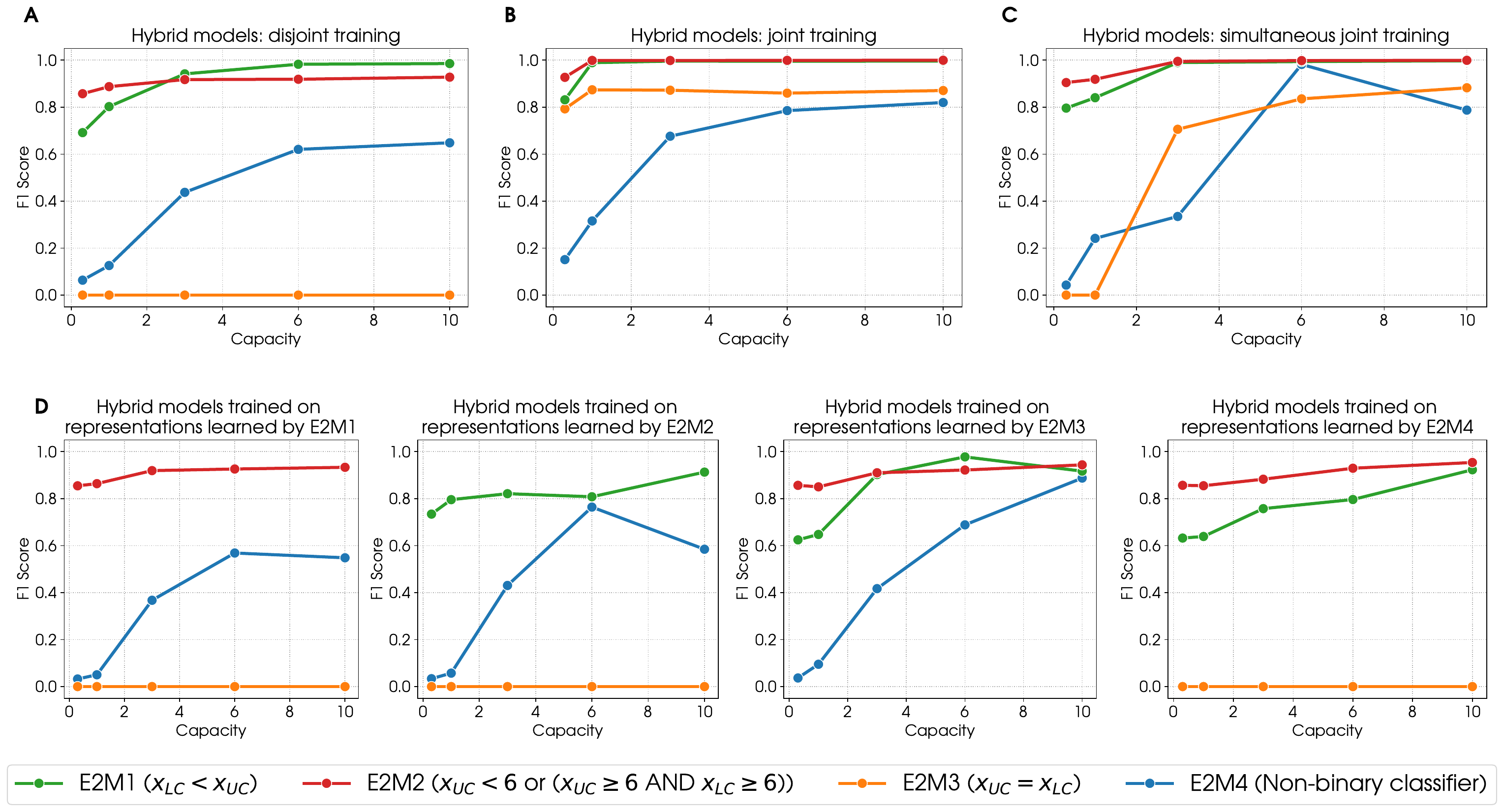}
\caption{Performance analysis. \rev{For ease of reading, a brief description of each model along with its label is reported in the legend.} \textbf{(A)} We first trained the baseline $\beta$-VAE model. Its frozen latent embeddings are then used to train separately 4 linear classifiers, one per task. This permits solving all tasks, except one. \textbf{(B)} We trained a $\beta$-VAE and classifier jointly and we repeated it for each task. In this case the classifier can modify the latent representation of the $\beta$-VAE. In fact, with respect to case A, the new representation is optimised for the given task and as result the network is able to solve the task1 and learn faster and better in the other tasks. \textbf{(C)} We trained only one $\beta$-VAE jointly with 4 classifiers (one per task). Such network is able to solve all the task simultaneously. \textbf{(D)} In this plot we asses the overspecialization of the networks trained in B. Consider the network trained on task 0. We use its latent representation (specialised for task0) to train 3 classifiers on the other tasks. What we can see is that task1 can never be solved using the representation learnt from an other task while this is not true for task 0, 2 and 3 (nevertheless, performance is worse than case B).}
\label{fig:f1_all}
\end{center}
\end{figure}

We next considered to what extent the baseline and hybrid models generalize to novel tasks, beyond those for which they were trained. For this, we assessed the performance of a frozen baseline $\beta$-VAE model, when linear classifiers are trained afterwards (Figure \ref{fig:f1_all}A). We found that using this sequential training regime permits solving all but one of the tasks, for which the embedding learned through unsupervised learning is not sufficient for downstream classification tasks, using a linear classifier. Furthermore, we found that the sequential training regime in which the $\beta$-VAE is trained before the classifiers (Figure~\ref{fig:f1_all}A) less efficient than the joint training regime of the hybrid models (E2M1 - E2M5), in which the $\beta$-VAE and the classifiers are trained jointly (Figure~\ref{fig:f1_all}B,C), across all the model / capacity combinations.

We next asked whether the classification tasks have the same of different capacity demands and what this implies for the latent representations developed by the models. We found that while almost all models reach a high F1 score, they require different capacities to do so. For example, in Figure \ref{fig:f1_all}B, the models  E2M1 and E2M3 reach a F1 score of 0.8 with a low capacity (0.3), whereas the model E2M4 reaches the same F1 score of 0.8 only with a high capacity (10). Since capacity influences reconstruction loss, two models having the same F1 score can have different reconstruction losses. For example, models E2M1 and E2M3 with a capacity of 0.3 (that is sufficient for a F1 score of 0.8) have a reconstruction loss of 0.15, whereas model E2M4 with a capacity of 10 has a reconstruction loss of 0.06 (Figure~\ref{fig:rec_loss_all}B). Interestingly, this results indicate that greater discrimination difficulty during learning induces better and more distinct input representations, which is in keeping with empirical findings \cite{ahissar1997task}.

Finally, we considered to what extent the hybrid models trained on one task generalize to other tasks when the representations of the $\beta$-VAE are frozen and novel classifiers are trained (Figure \ref{fig:f1_all}D). Comparing Figure~\ref{fig:f1_all}A with Figures~\ref{fig:f1_all}D permits appreciating that in some instances, training the $\beta$-VAE before the classifiers (Figure~\ref{fig:f1_all}A) leads to better performance compared to training jointly the $\beta$-VAE and one classifier and later train another classifier (Figures~\ref{fig:f1_all}D). However, this advantage is not systematic. Finally, comparing Figure~\ref{fig:f1_all}B and Figure~\ref{fig:f1_all}C permits appreciating that in many instances, jointly training multiple classifiers ((Figure~\ref{fig:f1_all}C)) achieves a performance that approximates models with one single classifier (Figure~\ref{fig:f1_all}B), especially at high capacity.

\subsubsection{Geometry of the representations of the models}

We next considered what representations emerge in the baseline $\beta$-VAE model that only uses unsupervised learning and to what extent the joint training of classification tasks (E2M1 - E2M5) distorts these representations.

%For these two panels, the datapoints are exactly the same across figures except for the color that varies according to the class the specific datapoints belongs to in the classification task). 

\paragraph{Baseline $\beta$-VAE model.}

\begin{figure}
\begin{center}
\includegraphics[width=\columnwidth]{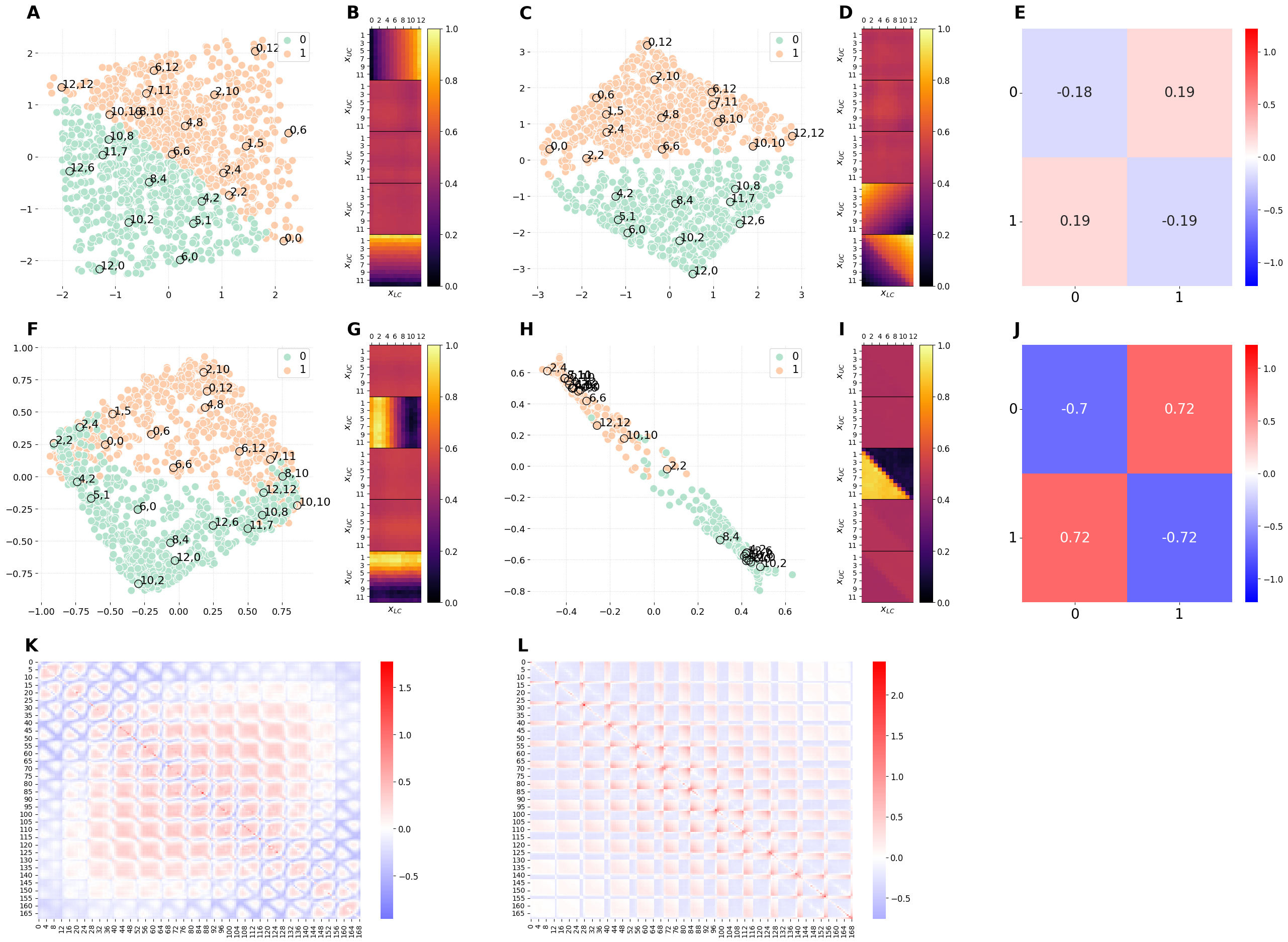}
\caption{
Comparison of the latent representations of the baseline model and the hybrid model E2M1. The E2M1 model is assigned a binary classification task, in which images with $x_U \leq x_L$ are labeled as $1$, $0$ otherwise.
\textbf{(A, F)} 2D projections of the 5D embeddings learned by the baseline model at high ($C_{\max}=10$ nats) and low ($C_{\max}=0.3$ nats) capacity, respectively.
\textbf{(C, H)} 2D projections of the 5D embeddings learned by the hybrid E2M1 model trained at high ($C_{\max}=10$ nats) and low ($C_{\max}=0.3$ nats) capacity, respectively. 
\textbf{(B, G)} Activation patterns of the 5 latent channels of the baseline model at high and low capacity, respectively. Each of the five heat-maps is computed as in sec. \ref{sec:neuron_activity}. 
\textbf{(D, I)} Activation patterns of the 5 latent channels of the E2M1 model at high and low capacity, respectively. Each heat-map is computed as in Section~\ref{sec:neuron_activity}. 
\textbf{(E, J)} Measure of distortions in the latent representations of the hybrid E2M1 model compared to the baseline model at high and low capacity, respectively. 
\textbf{(K)} Measure of distortions in the latent representation of the baseline model induced by the reduction of the encoding capacity from high to low.
\textbf{(L)} Measure of distortions in the latent representation of the E2M1 model induced by the reduction of the encoding capacity from high to low.
Distortion matrices are computed as described in Section~\ref{sec:compare_lr}.
}
\label{fig:E2M1}
\end{center}
\end{figure}

The representation learned by the baseline $\beta$-VAE model at high and low capacity is depicted in Figures~\ref{fig:E2M1}A and F, respectively. This is the same as Figure\ref{fig:E1M1}A and F, except that the dots have different colors. The colors of the dots indicates the class to which the image belongs in the classification task under analysis: green denotes class 0, while orange signifies class 1. While the baseline $\beta$-VAE model has no classification task, we color the dots anyway (here, with the classes used by the E2M1 model) for ease of comparison with the other models, see later.

\begin{figure}
\begin{center}
\includegraphics[width=\columnwidth]{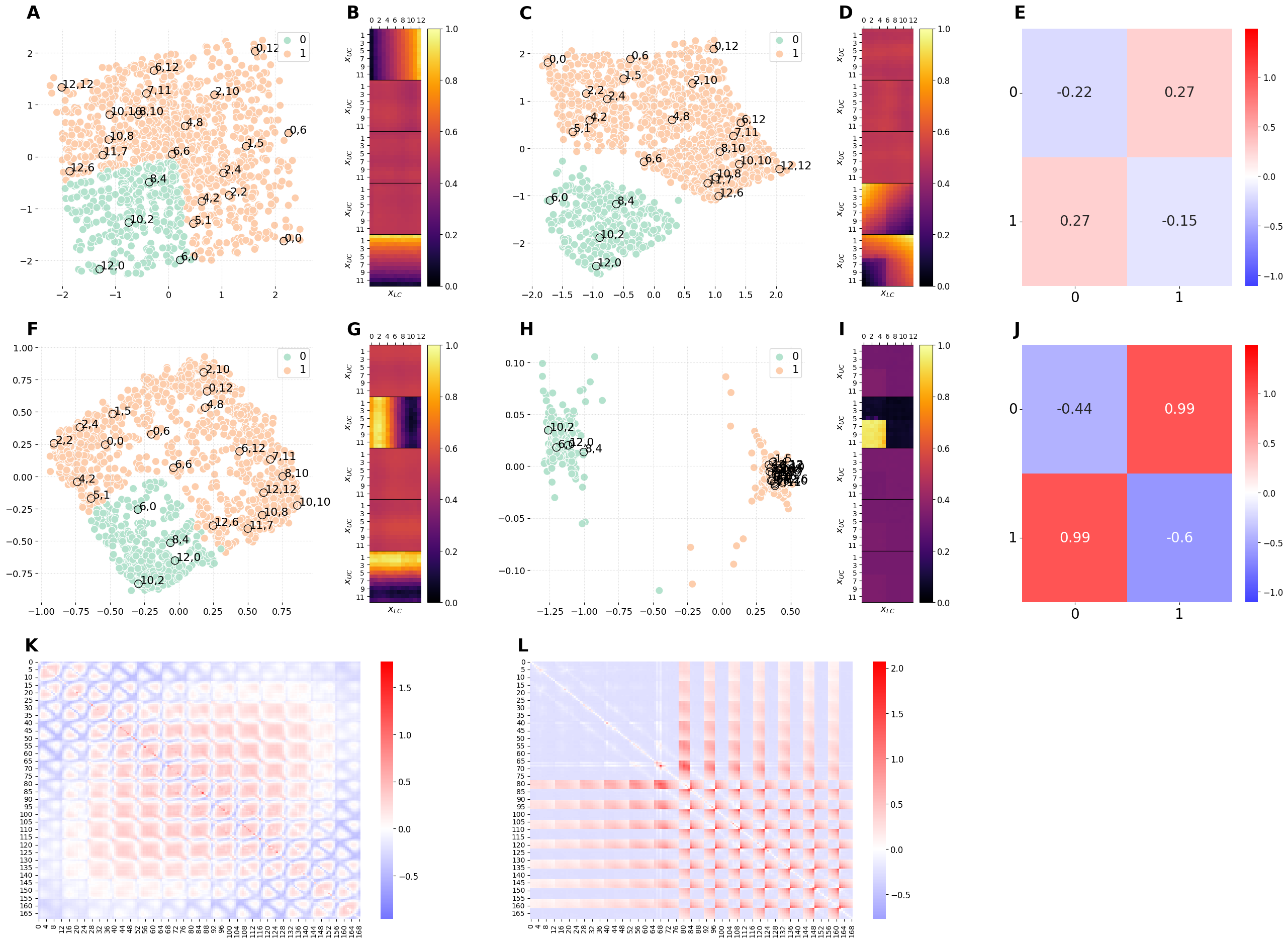}
\caption{Comparison of the latent representations of the baseline model and the hybrid model E2M2. The E2M2 model is assigned a binary classification task, in which images with $x_U <6$ (regardless of the value of $x_L$) OR $( x_U \geq 6$ AND $x_L \geq 6 )$ are labeled as $1$, $0$ otherwise.
\textbf{(A, F)} 2D projections of the 5D embeddings learned by the baseline model at high ($C_{\max}=10$ nats) and low ($C_{\max}=0.3$ nats) capacity, respectively.
\textbf{(C, H)} 2D projections of the 5D embeddings learned by the hybrid E2M2 model trained at high ($C_{\max}=10$ nats) and low ($C_{\max}=0.3$ nats) capacity, respectively. 
\textbf{(B, G)} Activation patterns of the 5 latent channels of the baseline model at high and low capacity, respectively. Each of the five heat-maps is computed as in sec. \ref{sec:neuron_activity}. 
\textbf{(D, I)} Activation patterns of the 5 latent channels of the E2M2 model at high and low capacity, respectively. Each heat-map is computed as in Section~\ref{sec:neuron_activity}. 
\textbf{(E, J)} Measure of distortions in the latent representations of the hybrid E2M2 model compared to the baseline model at high and low capacity, respectively. 
\textbf{(K)} Measure of distortions in the latent representation of the baseline model induced by the reduction of the encoding capacity from high to low.
\textbf{(L)} Measure of distortions in the latent representation of the E2M2 model induced by the reduction of the encoding capacity from high to low.
Distortion matrices are computed as described in Section~\ref{sec:compare_lr}.
}
\label{fig:E2M2}
\end{center}
\end{figure}

\begin{figure}
\begin{center}
\includegraphics[width=\columnwidth]{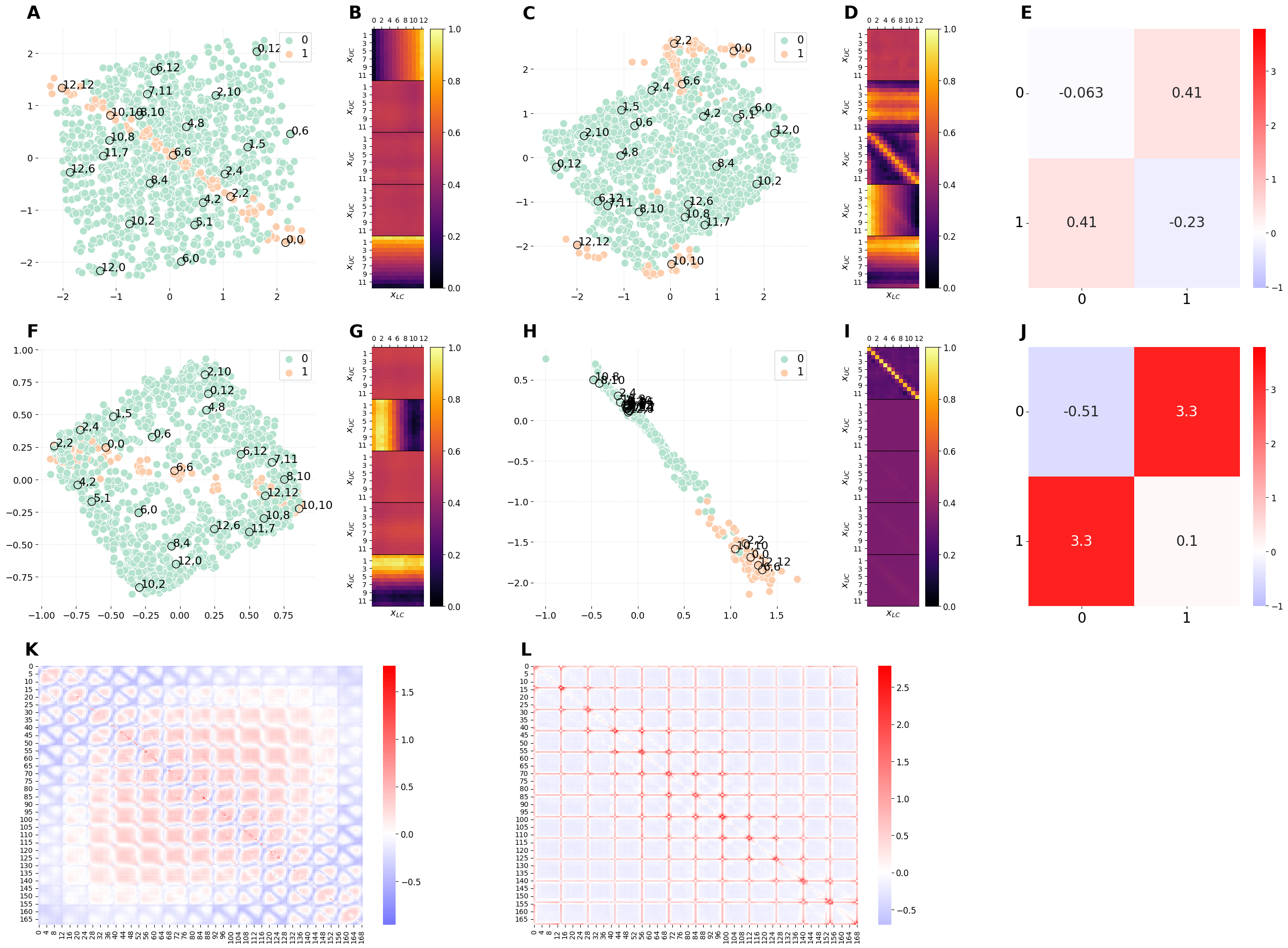}
\caption{
Comparison of the latent representations of the baseline model and the hybrid model E2M3. The E2M3 model is assigned a binary classification task, in which images with $x_U = x_L$ are labeled as $1$, $0$ otherwise.
\textbf{(A, F)} 2D projections of the 5D embeddings learned by the baseline model at high ($C_{\max}=10$ nats) and low ($C_{\max}=0.3$ nats) capacity, respectively.
\textbf{(C, H)} 2D projections of the 5D embeddings learned by the hybrid E2M3 model trained at high ($C_{\max}=10$ nats) and low ($C_{\max}=0.3$ nats) capacity, respectively. 
\textbf{(B, G)} Activation patterns of the 5 latent channels of the baseline model at high and low capacity, respectively. Each of the five heat-maps is computed as in sec. \ref{sec:neuron_activity}. 
\textbf{(D, I)} Activation patterns of the 5 latent channels of the E2M3 model at high and low capacity, respectively. Each heat-map is computed as in Section~\ref{sec:neuron_activity}. 
\textbf{(E, J)} Measure of distortions in the latent representations of the hybrid E2M3 model compared to the baseline model at high and low capacity, respectively. 
\textbf{(K)} Measure of distortions in the latent representation of the baseline model induced by the reduction of the encoding capacity from high to low.
\textbf{(L)} Measure of distortions in the latent representation of the E2M3 model induced by the reduction of the encoding capacity from high to low.
Distortion matrices are computed as described in Section~\ref{sec:compare_lr}.
}
\label{fig:E2M3}
\end{center}
\end{figure}

\begin{figure}
\begin{center}
\includegraphics[width=\columnwidth]{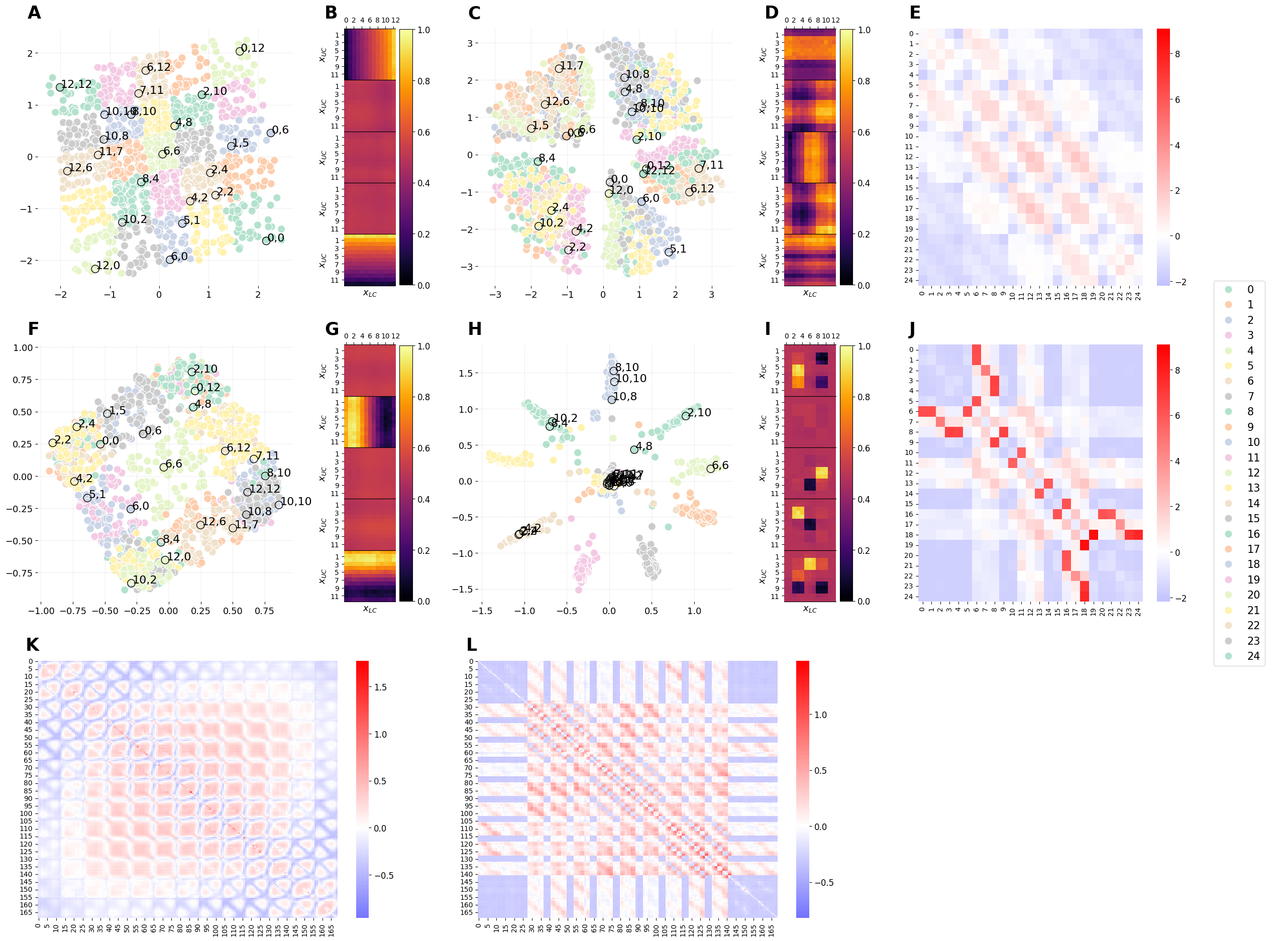}
\caption{
Comparison of the latent representations of the baseline model and the hybrid model E2M4. The E2M4 model is assigned a multiclass classification task. The possible positions of each corridor are grouped into 5 bins (as explained in Section~\ref{sec:exp2}) resulting in a partition of input images into 25 possible classes.
\textbf{(A, F)} 2D projections of the 5D embeddings learned by the baseline model at high ($C_{\max}=10$ nats) and low ($C_{\max}=0.3$ nats) capacity, respectively.
\textbf{(C, H)} 2D projections of the 5D embeddings learned by the hybrid E2M4 model trained at high ($C_{\max}=10$ nats) and low ($C_{\max}=0.3$ nats) capacity, respectively. 
\textbf{(B, G)} Activation patterns of the 5 latent channels of the baseline model at high and low capacity, respectively. Each of the five heat-maps is computed as in sec. \ref{sec:neuron_activity}. 
\textbf{(D, I)} Activation patterns of the 5 latent channels of the E2M4 model at high and low capacity, respectively. Each heat-map is computed as in Section~\ref{sec:neuron_activity}. 
\textbf{(E, J)} Measure of distortions in the latent representations of the hybrid E2M4 model compared to the baseline model at high and low capacity, respectively. 
\textbf{(K)} Measure of distortions in the latent representation of the baseline model induced by the reduction of the encoding capacity from high to low.
\textbf{(L)} Measure of distortions in the latent representation of the E2M4 model induced by the reduction of the encoding capacity from high to low.
Distortion matrices are computed as described in Section~\ref{sec:compare_lr}.
}
\label{fig:E2M4}
\end{center}
\end{figure}

\begin{figure}
\begin{center}
\includegraphics[width=\columnwidth]{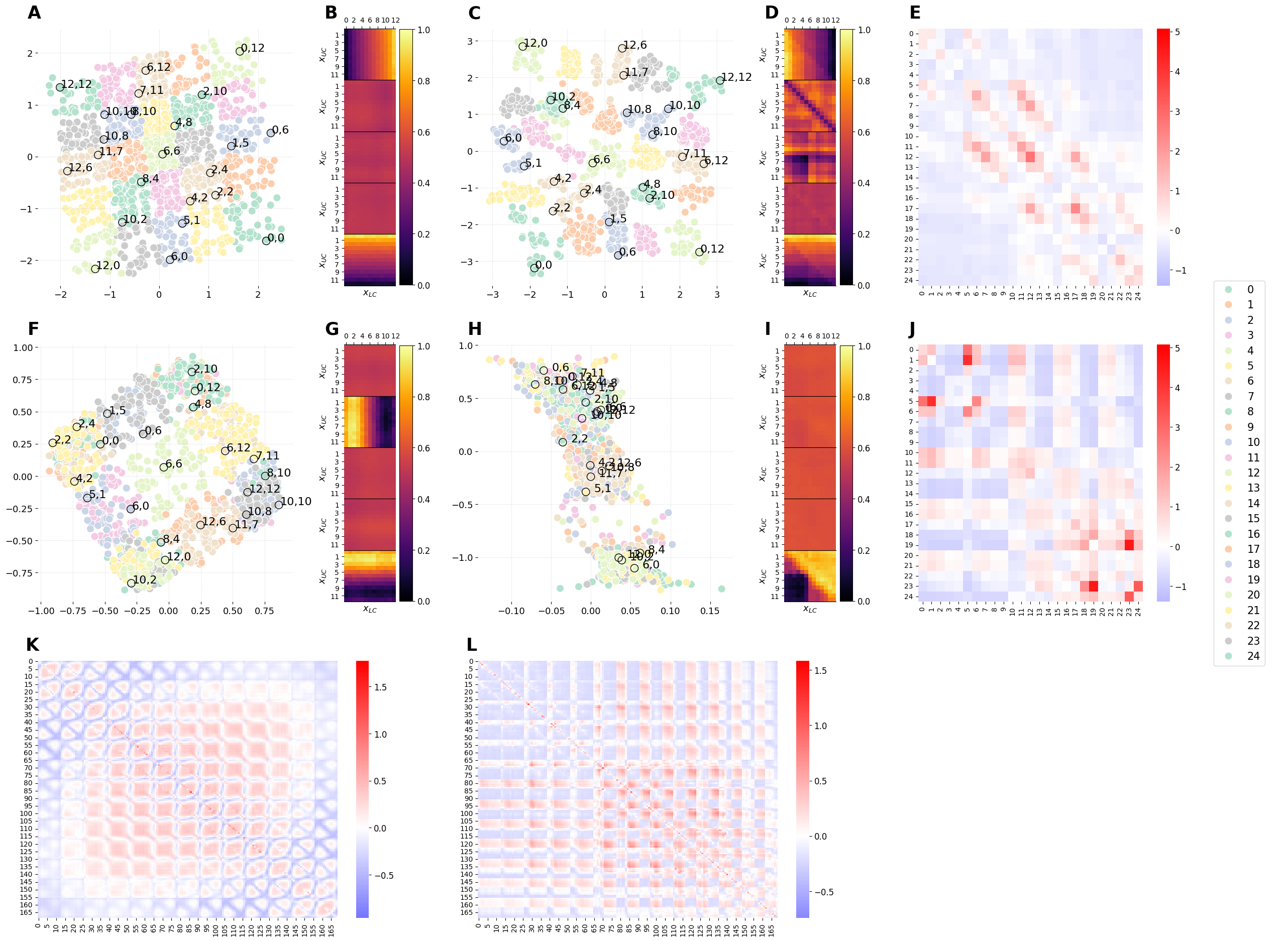}
\caption{
Comparison of the latent representations of the baseline model and the hybrid model E2M5. The E2M5 model is required to solve all the previous tasks simultaneously.
\textbf{(A, F)} 2D projections of the 5D embeddings learned by the baseline model at high ($C_{\max}=10$ nats) and low ($C_{\max}=0.3$ nats) capacity, respectively.
\textbf{(C, H)} 2D projections of the 5D embeddings learned by the hybrid E2M5 model trained at high ($C_{\max}=10$ nats) and low ($C_{\max}=0.3$ nats) capacity, respectively. 
\textbf{(B, G)} Activation patterns of the 5 latent channels of the baseline model at high and low capacity, respectively. Each of the five heat-maps is computed as in sec. \ref{sec:neuron_activity}. 
\textbf{(D, I)} Activation patterns of the 5 latent channels of the E2M5 model at high and low capacity, respectively. Each heat-map is computed as in Section~\ref{sec:neuron_activity}. 
\textbf{(E, J)} Measure of distortions in the latent representations of the hybrid E2M5 model compared to the baseline model at high and low capacity, respectively. 
\textbf{(K)} Measure of distortions in the latent representation of the baseline model induced by the reduction of the encoding capacity from high to low.
\textbf{(L)} Measure of distortions in the latent representation of the E2M5 model induced by the reduction of the encoding capacity from high to low.
Distortion matrices are computed as described in Section~\ref{sec:compare_lr}.
}
\label{fig:E2M5}
\end{center}
\end{figure}

%The latent representations learned by the hybrid model in E2M1, E2M2, E2M3, E2M4 and E2M5 are compared against the representation learned by the baseline model in Figures \ref{fig:E2M1}, \ref{fig:E2M2}, \ref{fig:E2M3}, \ref{fig:E2M4}, \ref{fig:E2M5} respectively. 

%In each figure we have 4 plots: plot A and C contain the latent representation learned by the baseline $\beta$-VAE when trained at high and low capacity respectively; plot B and D contain the latent representation learned by the hybrid network when trained at high and low capacity respectively. The plot of the latent representation for a given network is generated as described in section \ref{sec:compare_lr}. Note that plots A and C are exactly the same in all figures except for the color of the datapoints: the color of each datapoint is changed according to the class it belongs to for the given classification task.

\paragraph{Comparison between baseline and E2M1 models.}

Figure \ref{fig:E2M1} compares the latent representations learned by the baseline model at high capacity (Figure \ref{fig:E2M1}A) and low capacity (Figure \ref{fig:E2M1}F) with those learned by the hybrid model E2M1 at high capacity (Figures \ref{fig:E2M1}C) and low capacity (Figures \ref{fig:E2M1}H), respectively. These images help understand how the classification task distorts the latent representation of the model. 

At high capacity (Figure \ref{fig:E2M1}C), the representation retains the square-like shape of the baseline network (Figures \ref{fig:E2M1}A), but the datapoints of the two different categories are more segregated. This is evident by considering Figure \ref{fig:E2M1}E, which measures an average contraction of the distances between images belonging to the same class (in blue) and an average dilation for images of distinct classes (in red). Interestingly, as elucidated by Figure~\ref{fig:E2M1}D, the activation pattern of the latent channels of the E2M1 model completely changed and it is not aligned anymore with the two factors of variation in the dataset (i.e., there is no disentanglement) but rather it is a $45^{\circ}$-rotation of the baseline's latent representation: the axes of the new latent space correspond to the diagonals of the square in the baseline's latent space. In this way, the E2M1 model is able to solve the task by using only one latent channel and it uses the other latent factor to improve the fidelity of stimulus reconstruction.

When the E2M1 is trained at low capacity, the latent space reduces to a 1-dimensional space and only the axis allowing to solve the task is kept by the network (see Figure~\ref{fig:E2M1}I). As a consequence, the square-like shape is completely lost an all the datapoints collapse into two clusters, corresponding to the two categories that the model learned to discriminate (Figure \ref{fig:E2M1}H). In other words, the classification objective at low capacity induces an orthogonalization of representations, which is quantified in Figure \ref{fig:E2M1}J. This result is reminiscent of the empirical finding that learning visuomotor associations orthogonalizes population codes in the cortex \cite{failor2021visuomotor}. 

In sum, at high capacity, the classification task induces a weak form of orthogonalization of the latent space, which is manifest in a better separation of the points of the two classes compared to the baseline model. Rather, at low capacity, the classification task induces a strong orthogonalization of the latent state, forming two clusters organize around prototypes of the two classes. The differences between the distortions induced by the classifier with high and low capacity is further illustrated in Figure \ref{fig:E2M1}L, which shows a complex pattern of reorganization, with latent representations for the same class collapsing into separate clusters.

\paragraph{Comparison between baseline and E2M2 models.}

Figure \ref{fig:E2M2} compares the latent representations learned by the baseline model at high capacity (Figure \ref{fig:E2M2}A) and low capacity (Figure \ref{fig:E2M2}F) with those learned by the hybrid model E2M2 at high capacity (Figures \ref{fig:E2M2}C) and low capacity (Figures \ref{fig:E2M2}H), respectively. 

Due to the similarity with the task being solved by the E2M1 model (Figure~\ref{fig:E2M1}), the same observations hold true, confirming previous results. However, in this task, the numerosity of stimuli belonging to the two categories is different and this is reflected in the different sizes of their two corresponding subspaces (green and orange areas) in the latent representation.

\paragraph{Comparison between baseline and E2M3 models.}

Figure \ref{fig:E2M3} compares the latent representations learned by the baseline model at high capacity (Figure \ref{fig:E2M3}A) and low capacity (Figure \ref{fig:E2M3}F) with those learned by the hybrid model E2M3 at high capacity (Figures \ref{fig:E2M3}C) and low capacity (Figures \ref{fig:E2M3}H), respectively. 
As evident from Figure~\ref{fig:E2M3}D, at high capacity, the E2M3 network presents four active latent channels: one of them (the third from the top in Panel D) only activates when the input image has the two corridors aligned, while the remaining three active channels encode the corridor positions, similar to the baseline model. Thus, interestingly, the E2M3 model does not only encode task positions but also task-relevant information into a separate, almost-binary latent channel (the third one). The usefulness of this latent space representation becomes apparent when considering the nature of the task being solved: while the classification tasks of E2M1 and E2M2 models are linear, recognising corridors alignment represents a non-linear classification task. Since the classifier of the E2M3 model is linear, the classifier alone cannot solve the task on the basis of a representation like the one of the baseline model (see also Figures~\ref{fig:f1_all}A and D) and thus the network adapted the representation to explicitly encode alignment information enabling the linear classifier to correctly solve the task. This also explains the separation in the latent space between class 1 and class 0 images observable in Figure~\ref{fig:E2M3}C and quantified in Figure~\ref{fig:E2M3}E. This result shows that an explicit latent representation specialized for the task can emerge in complex classification tasks and coexist with latent representations that capture the generative factors and allow an accurate reconstruction of stimuli.

Interestingly, at low capacity, only the latent channel encoding task-relevant information is active, resulting in a latent space with two clusters (Figures~\ref{fig:E2M3}H and I). This latent space induces a strong separation of task-relevant inputs compared to the baseline model trained at low capacity, as it can be notice from a plot of their differences (Figures~\ref{fig:E2M3}J).

\paragraph{Comparison between baseline and E2M4 models.}

Figure \ref{fig:E2M4} compares the latent representations learned by the baseline model at high capacity (Figure \ref{fig:E2M4}A) and low capacity (Figure \ref{fig:E2M4}F) with those learned by the hybrid model E2M4 at high capacity (Figures \ref{fig:E2M4}C) and low capacity (Figures \ref{fig:E2M4}H), respectively. 

The channel activations illustrated in Figure \ref{fig:E2M4}D show that at high capacity, the E2M4 model exploits all the 5 latent channels to encode stimuli. Some channels show horizontal or vertical bands, indicating that they encode information about a single corridor. The presence of these channels indicates that the latent state is still to some extent disentangled; however, the fact that the bands are sharply separated indicates that the representation is distorted compared to the baseline model -- plausibly, to favor the performance of the model on the classification task. Furthermore, and interestingly, the channels coding for single corridors coexist with channels coding for both channels simultaneously, as evident from the fact that they show both horizontal and vertical bands. The coexistence of these different types of channels produce a significant rearrangement of the latent space compared to the baseline model, with different and separated subspaces allocated to the different stimulus classes (Figures \ref{fig:E2M4}C-E). 

When the E2M4 model is trained at low capacity, it does not have sufficient capacity to encode all the stimulus classes: indeed, only 9 of 25 classes are encoded in the latent representation, while remaining classes are collapsed into a single one (Figure \ref{fig:E2M4}H-J). This effect can be considered analogous to the 
\emph{specialization} effect observed with unbalanced datasets, in the sense that the model uses its limited resources to perform only partially its task. Note that the patterns shown by the latent channels of the E2M4 model trained at low capacity are more patchy compared to both the baseline model and the E2M4 model trained at high capacity (Figure \ref{fig:E2M4}I). This limited task representation explains the poor behavioral performance of the model (Figure~\ref{fig:rec_loss_all}B).

\paragraph{Comparison between baseline and E2M5 models.}

Figure \ref{fig:E2M5} compares the latent representations learned by the baseline model at high capacity (Figure \ref{fig:E2M5}A) and low capacity (Figure \ref{fig:E2M5}F) with those learned by the hybrid model E2M5 at high capacity (Figures \ref{fig:E2M5}C) and low capacity (Figures \ref{fig:E2M5}H), respectively. Note that while in the figures, the label colors reflect the same task as the E2M4 model, the E2M5 model is trained simultaneously on all the tasks solved by models E2M1 - E2M5. 

The latent channels of the E2M5 model trained at high capacity (Figure~\ref{fig:E2M5}D) reveal patterns that are compatible with a combination of those observed in the models E2M1 -- E2M4 -- which also explains the good performance of the E5M5 model across all the tasks (see Figure~\ref{fig:f1_all}C and Figure~\ref{fig:rec_loss_all}B). The first and last latent channels show a complex pattern that partly resembles tho one observed in the baseline model: specifically, the color is constant along one of the axis when considering the upper or lower triangular part only. Rather, the difference in color between the upper and lower triangular parts resembles the activation pattern of the E2M1 model. The second and third latent channels show patterns resembling the E2M3 and the E2M2 models, respectively. It is also possible to observe five color bands in each channel, compatible with the task solved by the E2M4 model. The resulting latent space is square-like, but the data points are further separated to accommodate all the tasks simultaneously (Figure~\ref{fig:E2M5}C). As an example, consider that the violet cluster containing the images with labels (5,1) and (6,0) is divided into two sub-clusters, to accommodate tasks solved by models E2M2 and E2M4. 

At low capacity, the E2M5 model learns to solve only 2 tasks (see Figure~\ref{fig:f1_all}C, showing that only the F1 scores of task associated to E2M1 and E2M2 models are high). This effect can be considered analogous to the 
\emph{specialization} effect observed with unbalanced datasets, in the sense that the model uses its limited resources to perform only partially its task. The latent channels of the E2M5 model resemble those of E2M1 and E2M2 models (Figure~\ref{fig:E2M5}I). This translates into a latent space with three clusters: the first cluster contains images having label 1 on both tasks, the second cluster contains images having label 0 on both tasks, and the third cluster contains images having different labels on the two tasks. %(note that in Figure~\ref{fig:E2M5}H, only two clusters are evident due to multidimensional scaling, but the third pattern can be appreciated by looking at Supplementary Figure~\ref{fig:E2M5_3D}).

\subsubsection{Summary of the results of Experiment 2}

Experiment 2 investigates the distortions in $\beta$-VAE embeddings resulting from varying model capacity and incorporating classification tasks. Five hybrid models, E2M1 to E2M5, are compared to a baseline $\beta$-VAE model, at different capacities. E2M1 to E2M4 are hybrid models combining $\beta$-VAE and one classifier, each trained on a specific classification task. E2M5 includes four classifiers trained simultaneously with $\beta$-VAE. 

Performance analysis reveals that as model capacity increases, reconstruction loss decreases across all models. However, hybrid models consistently exhibit higher reconstruction loss than the baseline model, due to the additional classification objective. Joint training of $\beta$-VAE and classifiers enables solving multiple tasks, with different capacities required for optimal performance in each task. Sequential training of $\beta$-VAE followed by classifiers yields less efficient performance compared to joint training, except in specific cases. Furthermore, joint training of multiple classifiers approaches the performance of single-classifier models, particularly at high capacity.

Regarding the geometry of representations, the classifiers induce task-specific distortions, as evident in comparisons between baseline and hybrid models. At high capacity, the distortions primarily improve segregation of class representations. Interestingly, explicit latent representations specialized for particular tasks can emerge and coexist in complex classification tasks with latent representations that capture the generative factors and allow an accurate reconstruction of stimuli. At low capacity, the distortions are more severe: the trained models show an orthogonalization of latent channels, with collapsed representations focused solely on task-relevant information -- which also implies a severe loss of stimulus reconstruction accuracy of performance. This phenomenon is consistent across models E2M1 to E2M4.

In E2M5, joint training enables the model to simultaneously solve multiple tasks, with latent representations accommodating all tasks. However, at low capacity, the model only learns to solve two tasks effectively, leading to a poor latent space with fewer distinct clusters compared to the number of tasks.

\section{Discussion}

When creating internal models of the world, the brain necessarily compresses information. Formal frameworks like rate-distortion theory provide a normative perspective explaining \emph{why} three different factors -- model capacity, data distributions and tasks -- distort latent representations. Here, instead, we ask \emph{how} these factors distort latent representations. 

To address this question, we employ a Beta Variational Autoencoder ($\beta$-VAE) \cite{bvae_2017}, which approximates rate-distortion theory, while allowing for the utilization of real-world stimuli such as images \cite{bates2020efficient,nagy2020optimal}. We use the $\beta$-VAE to conduct two computational experiments, in which we vary model capacity and data distributions (Experiment 1) and model capacity and classification task, while also augmenting the $\beta$-VAE with additional classification objectives (Experiment 2). 

We report six main findings. First, the $\beta$-VAE trained at high capacity successfully recovers efficient latent representations that \emph{disentangle} the two dimensions of variation of the stimuli (here, the positions of two corridors). However, lowering capacity distorts latent representations, inducing a \emph{prototypization} effect, which is evident in panels F. These compressed representations can be equally interpreted as \emph{abstract} representations with a categorical bias -- in the sense that they abstract away from sensory details \cite{bates2020efficient}.

Second, manipulating data distributions (i.e., training the model with unbalanced datasets) distorts latent representations, especially at low capacity. The resulting latent representations lose the disentanglement and show some \emph{specialization}: they faithfully represents the most common stimuli, but not the infrequent ones. Interestingly, this training regime also favors the development of higher dimensional latent representations, plausibly as a way to distinguish rare stimuli from frequent ones. Simultaneously lowering capacity and manipulating data distributions (i.e., using an unbalanced dataset) makes the prototype unbalanced, too. 

Third, introducing an additional objective -- classification accuracy -- penalizes reconstruction accuracy in hybrid models. Furthermore, classification accuracy produces an \emph{orthogonalization} effect -- and the increase of distance between latent representations for stimuli of different classes -- which is most evident for low capacity, but is also present for high capacity. This result is in line with the empirical finding that task training  orthogonalizes visual cortical population codes in rodents \cite{failor2021visuomotor}. Interestingly, in the hybrid models, the orthogonalization is not achieved by projecting information in a high dimensional space and then using linear readouts trained for specific computational tasks, as happens in reservoir computing and support vector machines \cite{maass2016searching}. Rather, it happens in the low dimensional space of VAEs. 

Fourth, certain hybrid models can specialize different latent channels for its two main objectives: reconstruction accuracy and classification. For example, in the case of model E2M3, an explicit latent channel specialized for the task coexists with latent channels that capture the generative factors and allow an accurate reconstruction of stimuli (Figure~\ref{fig:E2M3}D).

Fifth, the three kinds of distortions that we described -- \emph{prototypization}, \emph{orthogonalization} and \emph{specialization} -- can have compound effects. For example, the simultaneous presence of capacity limitations and task produces both prototypization and orthogonalization. This is most evident by looking at the latent representations of hybrid models trained at low capacity (e.g., Figures~\ref{fig:E2M2}H and \ref{fig:E2M1}H), which show a separation of two orthogonal prototypes. Interestingly, these are not the same prototypes that appear in the baseline model trained at low capacity (Figures~\ref{fig:E2M2}F and \ref{fig:E2M1}), but are rather amplifications of the separation between classes already observed when the hybrid models are trained at high capacity (Figure~\ref{fig:E2M1}C and Figure~\ref{fig:E2M2}C). 

Sixth, task difficulty significantly affects the specificity of encoding and perceptual learning. Specifically, greater discrimination difficulty during perceptual learning induces more distinct representations of individual images, which is in keeping with empirical findings \cite{ahissar1997task}. This is because easy tasks can be addressed with a greater level of compression.

Taken together, these results illustrate that three main classes of distortions of latent representations -- prototypization, specialization, orthogonalization -- emerge as signatures of information compression, under constraints on capacity, data distributions and tasks. These distortions can coexists, giving rise to a rich landscape of latent spaces, whose geometry could differ significantly across generative models subject to different constraints.

These results could be potentially useful to interpret neural data. Various empirical studies reported that the geometry of latent representations in neuronal populations is distorted, as in the case of reward-induced compression of spatial representations of mazes in the hippocampus and orbitofrontal cortex \cite{juechems2021optimal} or frequency-induced biases of working memory representations \cite{panichello2019error}. These and other empirical results can be interpreted in the light of the necessity for the brain to efficiently compress information -- keeping with rate-distortion theory \cite{shannon1959coding}. Interestingly, the brain is able to flexibly adapt the type and level of compression to task demands, rather than employing a fixed compression level. For example, the degree of specificity of internal representations along the visual pathway depends on the difficulty of the training conditions, with more specific representations emerging when learning difficult tasks \cite{ahissar1997task}. Furthermore, the prefrontal cortex can rapidly remap the representation of stimuli according to their usefulness \cite{castegnetti7usefulness}. It remains to be established in future studies whether these empirical findings can be characterized formally using rate-distortion theory.

%\textcolor{red}{Nota su capacità adattiva del cervello (model selection on the fly). Capacità maggiore implica codice piu preciso e più memoria}

Furthermore, future studies might use the computational framework developed here for the ambitious objective to reverse-engineer the brain's training objective from the geometry of latent representations of neuronal populations. Some researchers have argued that the ventral visual stream pursues the goal of perceptual categorization \cite{yamins2016using} and others that it pursues efficient data compression or efficient coding \cite{barlow1961possible}. Our analysis shows that these two goals induce different types of distortion; furthermore, it shows that the two goals are not mutually exclusive, but can be pursued in parallel. The relative importance of these two goals for the ventral visual stream (or other brain structures) could be in principle identified, by analyzing the geometry of latent representations.

This work has some limitations that could be addressed in future work. First, while our method assumes that reconstruction is part of the loss function (Figure~\ref{fig:intro1}B), other methods, such as the information bottleneck \cite{tishby2000information} dispense from reconstruction. It remains to be assessed whether the results obtained here generalize to these methods. Second, for the sake of simplicity, we adopted a simple training scheme for the hybrid models of Experiment 2, in which the $\beta$-VAE model and the classifier are trained together (with the exception of the results shown in Figure \ref{fig:f1_all}A, in which the classifier is trained after the $\beta$-VAE was fully trained and frozen). An alternative possibility is adopting a more sequential training regime, in which the classifier is trained after the $\beta$-VAE has developed stable latent representations. Preliminary investigations shows that this sequential training regime produces additional changes to the geometry of the representation, but these changes need to be investigated in more detail. Third, future studies might consider more sophisticated generative architectures (e.g., hierarchical or conditional $\beta$-VAEs) , which permit addressing the trade-offs between the capacity and accuracy afforded by different models \cite{akuzawa2021conditional} as well as more sophisticated methods to quantify how well latent representations support classification objectives \cite{cohen2019separability}. Fourth, while we have focused on relatively simple measures of the geometry of latent representations of $\beta$-VAEs and their distortions, future studies might consider more advanced analysis methods, which are based on interventions \cite{leeb2022exploring}. \rev{Finally, we focused on a specific dataset -- the Corridors dataset -- which we created specifically to study the distortions of generative models. We opted to create a new dataset instead of using standard datasets like MNIST \cite{deng2012mnist} because the lack of clearly defined orthogonal factors of variation would make it difficult to interpret the effects of rate-distortion trade-offs. While other less standard existing datasets like dSprites \cite{dsprites17} do feature orthogonal factors, they usually include a large number of independent dimensions of variation, significantly increasing the complexity of the analysis. By contrast, the Corridors dataset was designed to contain only two orthogonal generative factors, allowing us to systematically study how capacity constraints, data distributions, and task requirements distort latent representations. We expect our results to generalize to other datasets with a larger number of disentangled dimensions. Furthermore, we expect our methods to be particularly useful to study neural representations in biological systems and their distortions, which have been demonstrated in previous empirical studies \cite{muhle2023goal,castegnetti7usefulness,failor2021visuomotor}. However, the generalization of our methods to other datasets and their application to address biological systems remain to be addressed in future research.}

\section{Methods}

\subsection{Formal framework: Rate-distortion theory}

Rate-distortion theory (RDT) \cite{shannon1959coding} provides a normative framework to understand information compression and memory optimization in artificial and biological systems (Figure \ref{fig:intro1}). RDT is a fundamental aspect of information theory, forming the basis for lossy data compression techniques. While information theory primarily deals with minimal information loss in transmitting messages over noisy channels, RDT addresses scenarios where the channel capacity is insufficient for transmitting the entire message. It describes the optimal trade-off between fidelity and the constraints imposed by the channel capacity \cite{tishby2000information,tishby2010information2,lancia2023humans,pezzulo2024neural,polani2009information,rubin2012trading}.

According to RDT, efficient representations maximize memory performance, subject to a capacity limit or encoding cost. There is therefore a trade-off between available resources or capacity of the system and the distortion. This trade-off is schematically illustrated in Figure \ref{fig:intro1}A, for the reconstruction of one of the images of the Corridors dataset (input image). The figure shows that when the capacity or rate ($R_q$) is high, the distortion ($D_q$) is low and the reconstructed image is very similar to the input image. Conversely, when the capacity or rate ($R_q$) is low, the distortion ($D_q$) is high and the reconstructed image is a low-fidelity version of the input image. 

Figure~\ref{fig:intro1}B specifies formally the rate-distortion framework. Given an encoder model $q(\boldsymbol{z} | \boldsymbol{x})$, the expected amount of distortion in the reconstruction of the input $x$ from the embedding $z$ given by $q$ is specified by the distortion function $D_q$.
The rate function $R_q$ specifies the available resources or capacity for the encoder $q$ to retain information about the input $x$. In rate-distortion theory, the objective is to identify an optimal encoder model $q^*$ that minimizes the distortion $D_q$ while operating within a limited encoding capacity $R_q$, constrained by a constant $C$. 
\rev{
The Lagrangian $\mathcal{L}$ of this constrained optimization problem is:
$$
\mathcal{L} = D_q - \beta \left( R_q - C \right).
$$
} 
The Lagrange multiplier $\beta$ sets the trade-off between rate and distortion as elucidated in Figure~\ref{fig:intro1}A.

Rate-distortion theory characterizes formally the distortions due to available resources, frequency of the input, and their utility to achieve goals. Figure \ref{fig:intro1}D illustrates schematically how the available resources (rate) and the frequency of the input interact during image reconstruction. With a high rate, the reconstruction is accurate, regardless of the frequency of the input. With lower rates, however, the reconstruction depends on the frequency of the data in the dataset. If the input is very frequent in the dataset, it will be reconstructed as a blurry version of the original image. If the input is rare, it will be reconstructed as a blurry version of the most frequent type of input (or prototype) in the dataset (Figure \ref{fig:intro1}D). 

Figures \ref{fig:intro1}E illustrates schematically how the available resources (rate) and the utility of the input interact during the reconstruction of two images. With high rate, the reconstruction if always accurate. However, with low rate, reconstruction accuracy depends on utility. If the differences between the two images are relevant (i.e., the images belong to two different classes of a binary classification task), the images are reconstructed as blurry versions of the inputs, with their differences preserved, see the top-left part of Figures \ref{fig:intro1}E. Rather, if the differences between the two images are not relevant (i.e., the images belong to the same class of a binary classification task), they are reconstructed in very similar ways and their differences are no more evident, see the bottom-left part of Figures \ref{fig:intro1}E. This example shows that at low capacity, the system maps all the input images into two noisy prototypes, one for each classification task.

\subsection{Computational Models}
\label{sec:models}

While exact methods are effective for discovering optimal solutions of RDT in low-dimensional input spaces, addressing higher-dimensional data like the images used in this study requires the use of approximate methods, like deep neural networks. In our study, we employed a Beta Variational Autoencoder ($\beta$-VAE) \cite{bvae_2017} -- a generative deep learning method inspired by variational inference methods in statistics -- to compress the original input space into a lower-dimensional latent space, to study what latent representations of data emerge under different task distributions and goals. The $\beta$-VAE is considered an effective approximation of RDT for high-dimensional data, such as images \cite{bates2020efficient,nagy2020optimal}. This is because the loss function of the $\beta$-VAE has a strong formal relation to RDT (Figure \ref{fig:intro1}B). 
\rev{The loss function of a $\beta$-VAE can be written as follows \cite{bvae_2018}:
%$$
\begin{equation}\label{eq:bvae_loss}
\mathcal{L}(\theta, \phi ; \boldsymbol{x}, \boldsymbol{z}, \beta)=
\mathbb{E}_{q_{\phi}(\boldsymbol{z} \mid \boldsymbol{x})}\left[\log p_{\theta}(\boldsymbol{x} \mid \boldsymbol{z})\right] - 
\beta D_{KL}\left(q_{\phi}(\boldsymbol{z} \mid \boldsymbol{x}) \| p(\boldsymbol{z})\right).
\end{equation}
%$$
It} comprises two terms: a reconstruction loss and a regularization term. The reconstruction loss $\mathbb{E}_{q_{\phi}(\boldsymbol{z} \mid \boldsymbol{x})}\left[\log p_{\theta}(\boldsymbol{x} \mid \boldsymbol{z})\right]$ quantifies the discrepancy between the original input data $x$ and its reconstruction $\hat{x}$ obtained from the latent embedding $z$. This loss term reflects the distortion function $D_q$ in rate-distortion theory, as it measures the fidelity of the reconstructed data compared to the original input. On the other hand, the regularization term $D_{K L}\left(q_{\phi}(\boldsymbol{z} \mid \boldsymbol{x}) \| p(\boldsymbol{z})\right)
$ \rev{- the Kullback–Leibler divergence -} penalizes the deviation of the learned latent distribution $q_{\phi}(z|x)$ from a predefined prior distribution $p(\boldsymbol{z})$, typically Gaussian. This regularization term can be associated with the rate function $R_q$, which delineates the capacity of the encoder to capture relevant information about the input $x$.

We adopt the $\beta$-VAE in both Experiments 1 and 2. However, while in Experiment 1 the $\beta$-VAE is \rev{
trained with the loss in Equation \ref{eq:bvae_loss}, in Experiment 2 the $\beta$-VAE is combined with one (or more) classifier(s),i.e., it is trained with a different loss:

\begin{equation}\label{eq:bvae+clf_loss}
    \mathcal{L}(\theta, \phi ; \boldsymbol{x}, \boldsymbol{z}, \beta)=
\mathbb{E}_{q_{\phi}(\boldsymbol{z} \mid \boldsymbol{x})}\left[\log p_{\theta}(\boldsymbol{x} \mid \boldsymbol{z})\right] - 
\beta D_{KL}\left(q_{\phi}(\boldsymbol{z} \mid \boldsymbol{x}) \| p(\boldsymbol{z})\right) + 
\sum_{i=1}^{n} \mathcal{L}_{\operatorname{clf}}^{(i)} \left( \boldsymbol{z}, \boldsymbol{y}^{(i)} \right),
\end{equation}
where the last term represents the sum of the losses $\mathcal{L}_{\operatorname{clf}}^{(i)}$ of the classifiers that are jointly trained with the $\beta$-VAE. It is possible to have only one classifier ($n=1$), as for models E2M1-E2M4, or more than one ($n>1$), as for model E2M5 which features $n=4$. The $i$-th classifier receives as input the latent vector $\boldsymbol{z}$ from the $\beta$-VAE and the label $\boldsymbol{y}^{(i)}$ associated to $\boldsymbol{z}$ in the $i$-th task. This joint training procedure allows us to study the distortion induced by the classification task on the latent representation learnt by the $\beta$-VAE models. Depending on the specific task, the classifier can be linear or non-linear. The linear classifier is a feedforward neural network with no activations and a single hidden layer with 1024 units. The non-linear classifier is a feedforward neural network with leaky ReLU activations and a single hidden layer with 1500 units. 
This difference between Experiment 1 and 2 is schematically reported in Figure \ref{fig:intro1}B and C by means of the yellow dashed box.
}

%stand-alone (see Figure \ref{fig:intro1}C, without the part shown in the yellow box) in Experiment 2 it is combined with one (or more) classifier(s) (see Figure \ref{fig:intro1}C, with the part shown in the yellow box).

The $\beta$-VAE used for both Experiments 1 and 2 comprises an encoder block and a decoder block. The encoder block consists of 3 convolutional layer, each followed by a leaky ReLU activation. Similarly, the decoder block consists of 3 transposed convolutional layers, also followed by a leaky ReLU activation function, except for the last layer having a sigmoid activation function to output grayscale images. 

\rev{Each model has been trained 5 times,  each time with a different value of $C_{\operatorname{max}}$, the maximum encoding capacity. The encoding capacity C represents the amount of information each latent dimension of the VAE can encode about the input and it depends on the  KL divergence between the approximate posterior $q_{\phi}(\boldsymbol{z} \mid \boldsymbol{x})$ and the prior $p(\boldsymbol{z})$. Following \cite{bvae_2018}, and re-writing the loss in Equation \ref{eq:bvae_loss} as:
\begin{equation}\label{eq:bvae_loss_C}
\mathcal{L}(\theta, \phi ; \boldsymbol{x}, \boldsymbol{z}, \beta)=
\mathbb{E}_{q_{\phi}(\boldsymbol{z} \mid \boldsymbol{x})}\left[\log p_{\theta}(\boldsymbol{x} \mid \boldsymbol{z})\right] - 
\gamma \left| D_{KL}\left(q_{\phi}(\boldsymbol{z} \mid \boldsymbol{x}) \| p(\boldsymbol{z})\right) - C\right|
\end{equation}
it is possible to control the encoding capacity C of the network by penalizing with a factor $\gamma$ any deviation of the KL divergence from the required capacity C. During training, C is gradually increased from $0$ to a target value $C_{\operatorname{max}}$ to favor the disentanglement of latent representation \cite{bvae_2018}. In our experiments, possible values for $C_{\operatorname{max}}$ are  0.1, 1, 3, 6 and 10 nats. 

The latent dimension is fixed to 5 for all the models, since there are only two generative factors of variation in the data. We added three extra dimensions to see whether the models are able to use them efficiently or not.

All the models are trained with Adam optimizer and a learning rate of $5\cdot10^{-5}$. 
}

\subsection{Comparing latent representations (or embeddings) learned by the different models}\label{sec:compare_lr}

Our goal is to compare the geometry of the latent representations (or embeddings) learned by different models in our experiments. Since all models present a 5-dimensional latent space, we employ Multidimensional Scaling (MDS) \cite{borg2005modern} as a dimensionality reduction technique to render 2D representations of the latent space structure.

To execute MDS, we input a dissimilarity matrix derived from pairwise Euclidean distances between data points in the high-dimensional space. For each trained model, we generate 2D projections of the latent space by applying MDS to the pairwise Euclidean distance matrix computed from the 5-dimensional vectors associated with each image via the encoder block of the respective network. Then, we compare the 2D embeddings learned by the different models considered (see panels A, C, F and H of figures \ref{fig:E1M1}, \ref{fig:E1M2}, \ref{fig:E2M1}, \ref{fig:E2M2}, \ref{fig:E2M3}, \ref{fig:E2M4} and \ref{fig:E2M5}); see also Supplementary Figures \ref{fig:E1M1_3D} - \ref{fig:E2M5_3D} for 3D plots of the embeddings.

\paragraph{Class distortion matrix and Item distortion matrix}

Given two models, $M_1$ and $M_2$, a way to quantitatively measure how much the latent space of model $M_2$ is distorted with respect to that of model $M_1$ is to measure, for each pair of images $u$ and $v$, how far the distance between $M_2(u)$ and $M_2(v)$ deviates from the distance between $M_1(u)$ and $M_1(v)$, where $M_{\bullet}(u)$ denotes the embedding of image $u$ according to model $M_{\bullet}$. More specifically, we compute the ratios

\begin{equation}\label{eq:rho}
\rho (u,v) = \dfrac {d \left( M_2(u), M_2(v) \right)} {d \left( M_1(u), M_1(v) \right)}, \qquad \forall u \neq v.
\end{equation}
for all pairs of images employing the Euclidean distance as $d(\cdot, \cdot)$. When $\rho > 1$, the two images are farther apart in the latent space of $M_2$ with respect to $M_1$. When $\rho < 1$, the two images are closer in the latent space of $M_2$ with respect to $M_1$. When $\rho = 1$, the two images have the same distance in the latent space of both models. This allows us to measure the magnitude of dilation or compression of the embeddings of a model with respect to those of an other model.

Panel K of Figures \ref{fig:E1M1}, \ref{fig:E1M2}, \ref{fig:E2M1}, \ref{fig:E2M2}, \ref{fig:E2M3}, \ref{fig:E2M4} and \ref{fig:E2M5} reports the distortion matrix arising from a reduction of the encoding capacity of the baseline model. In particular, we consider the baseline model at high capacity as $M_1$ and the same model at low capacity as $M2$. To compute such matrix, we first assign a label to each image according to the position of the corridors: for a image with corridors in position $(i,j)$, we assign the label $l=13*i + j$. In total we have $169$ distinct labels. Each element $(a,b)$ of the matrix then represents the average value of $\rho$ computed on all the couples of images in which the first element is an image with label $a$ and the second element is a image with label $b$. This tells us how much, on average, images with labels $a$ are compressed or dilated with respect to images with label $b$ when passing from the latent space of model $M_1$ to that of model $M_2$. 
Similarly for panel L, except that we use one of the models among  E1M1, E1M2, E2M1, E2M2, E2M3, E2M4, E2M5 instead of the baseline model.

Panel E of Figures \ref{fig:E1M1}, \ref{fig:E1M2}, \ref{fig:E2M1}, \ref{fig:E2M2}, \ref{fig:E2M3}, \ref{fig:E2M4} and \ref{fig:E2M5} reports the distortion matrix arising when having an unbalanced dataset or a task assigned to the network at high capacity. In particular, we consider the baseline model at high capacity as $M_1$ and a "hybrid" model at high capacity as $M_2$. In this case, we group images into classes depending on the experiment. In experiment 1, for example, it is possible to divide images according to their frequency of appearance in the dataset (high and low frequency) and thus we have two classes ($0$ and $1$ labels). In experiment 2, it is possible to divide images into classes according to the specific classification task being solved by the network. The number of classes will hence depend on that task. By aggregating images according to the class they belong to, we can measure the average distortion for couple of images within the same class or belonging to distinct classes. This tells us how much, on average, images of one class are compressed/dilated with respect to images of an other class when passing from the latent space of model $M_1$ to that of model $M_2$.
Similarly for panel F, the models involved are the same of panel E but they are trained at low capacity.

\subsection{Comparing neurons' activity of the different models}\label{sec:neuron_activity}

Another way to qualitatively assess the distortion induced by a task (or stimuli statistics) onto the embedding space is to visualize the activation patterns of the individual latent channels of the 5D latent space when specific stimuli are passed as input to the model. For each latent channel, we create a $13x13$ matrix whose $(i,j)$ element is the average value of that channel when a image with corridors in position $(i,j)$ is passed as input, where the average is computed on all the images on the test set. For each model we thus obtain 5 matrices, one for each dimension of the latent space, and we plot them as heat maps (panels B, D, G and I of figures \ref{fig:E1M1}, \ref{fig:E1M2}, \ref{fig:E2M1}, \ref{fig:E2M2}, \ref{fig:E2M3}, \ref{fig:E2M4} and \ref{fig:E2M5}). This approach facilitates the identification of significant differences between models, such as variations in the number of active channels within the representation or their distinct activation patterns.

\section*{Acknowledgements}

This research received funding from the European Union’s Horizon 2020 Framework Programme for Research and Innovation under the Specific Grant Agreement No. 952215 (TAILOR) to GP; the European Research Council under the Grant Agreement No. 820213 (ThinkAhead) to GP, the Italian National Recovery and Resilience Plan (NRRP), M4C2, funded by the European Union – NextGenerationEU (Project IR0000011, CUP B51E22000150006, “EBRAINS-Italy”; Project PE0000013, “FAIR”; Project PE0000006, “MNESYS”) to GP, and the PRIN PNRR P20224FESY to GP. The GEFORCE Quadro RTX6000 and Titan GPU cards used for this research were donated by the NVIDIA Corporation. The funders had no role in study design, data collection and analysis, decision to publish, or preparation of the manuscript.

\bibliographystyle{vancouver}
\bibliography{src/references}

\newpage
\appendix

\section{Supplementary Materials}

\counterwithin{figure}{section}
\setcounter{figure}{0}
\renewcommand{\thefigure}{S\arabic{figure}}
\renewcommand{\theHfigure}{S\arabic{figure}}

%\subsection{Supplementary figures: 3D Plots}

%\newpage

\begin{figure}[h!]
\begin{center}
\includegraphics[width=\columnwidth]{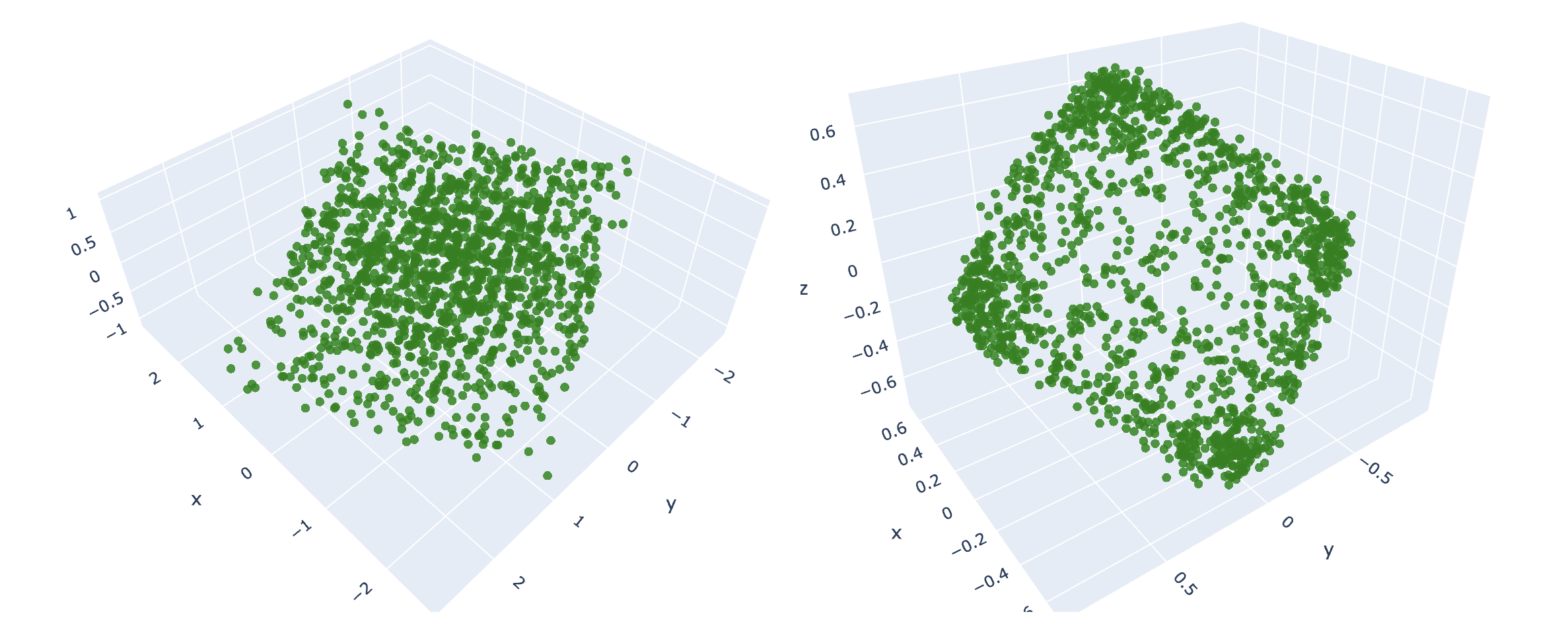}
\caption{
Comparison of the latent representations of the baseline model, trained at high capacity (left) and low capacity (right), in three dimensions.
}
\label{fig:baseline_3D}
\end{center}
\end{figure}

%\newpage

\begin{figure}[h!]
\begin{center}
\includegraphics[width=\columnwidth]{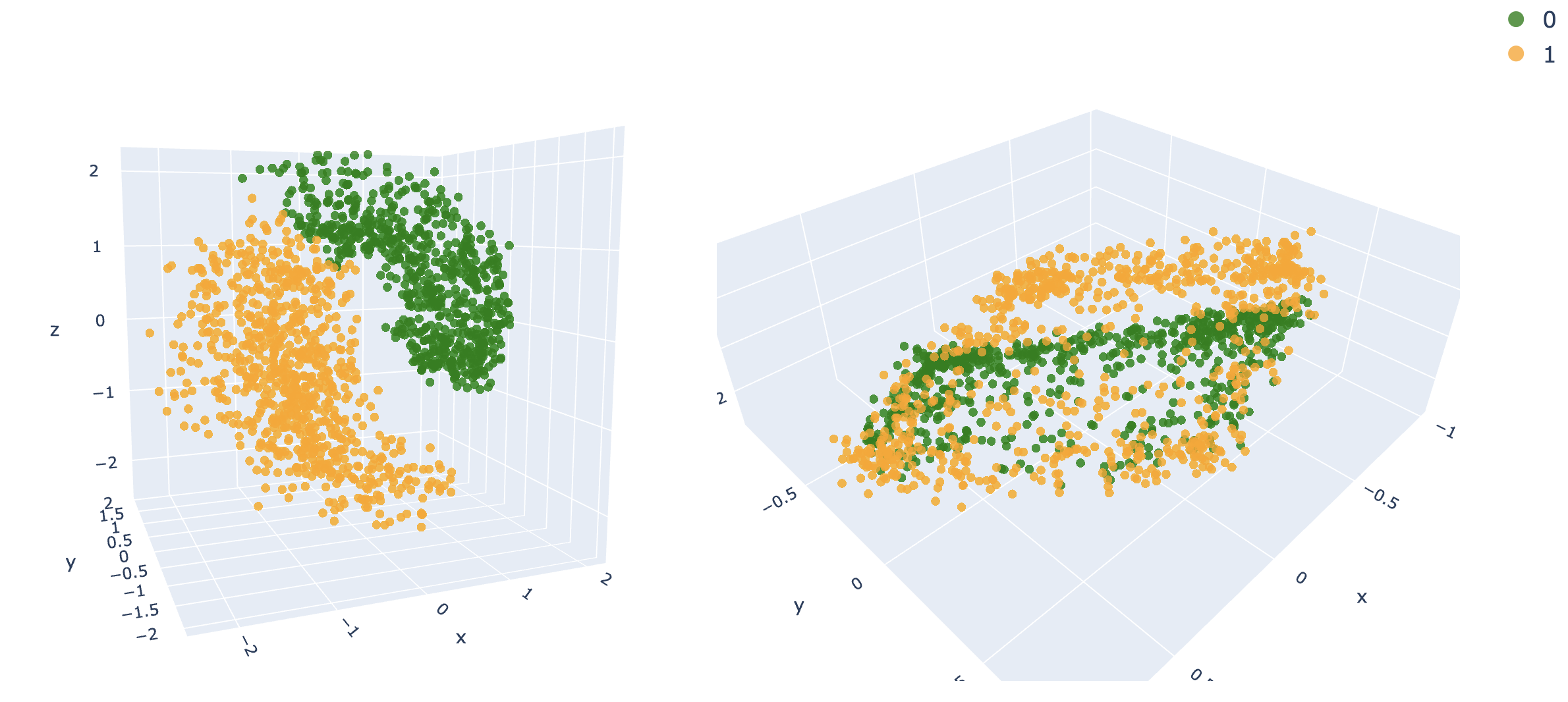}
\caption{
Comparison of the latent representations of the model E1M1, trained at high capacity (left) and low capacity (right), in three dimensions.
}
\label{fig:E1M1_3D}
\end{center}
\end{figure}

%\newpage

\begin{figure}[h!]
\begin{center}
\includegraphics[width=\columnwidth]{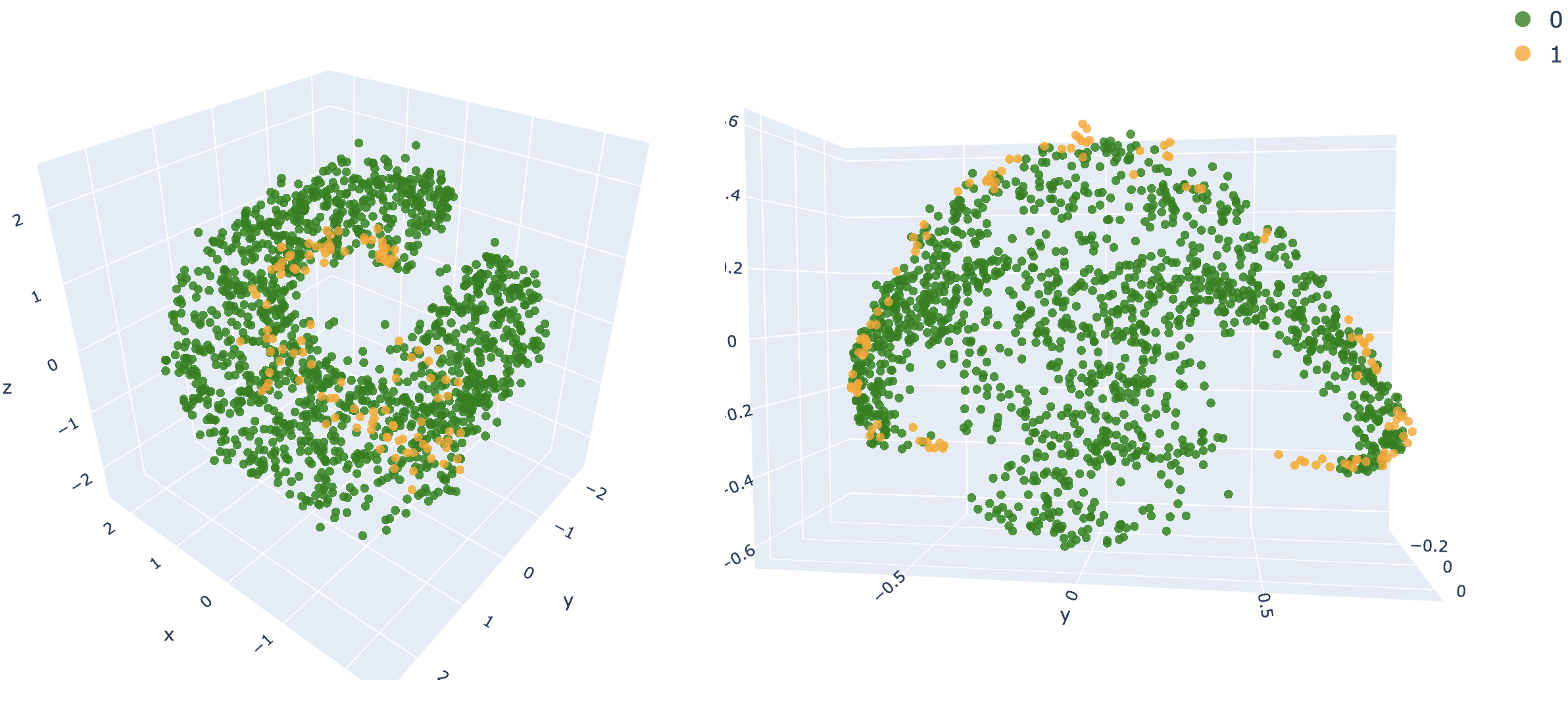}
\caption{
Comparison of the latent representations of the model E1M2, trained at high capacity (left) and low
capacity (right), in three dimensions.
}
\label{fig:E1M2_3D}
\end{center}
\end{figure}

%\newpage

\begin{figure}[h!]
\begin{center}
\includegraphics[width=\columnwidth]{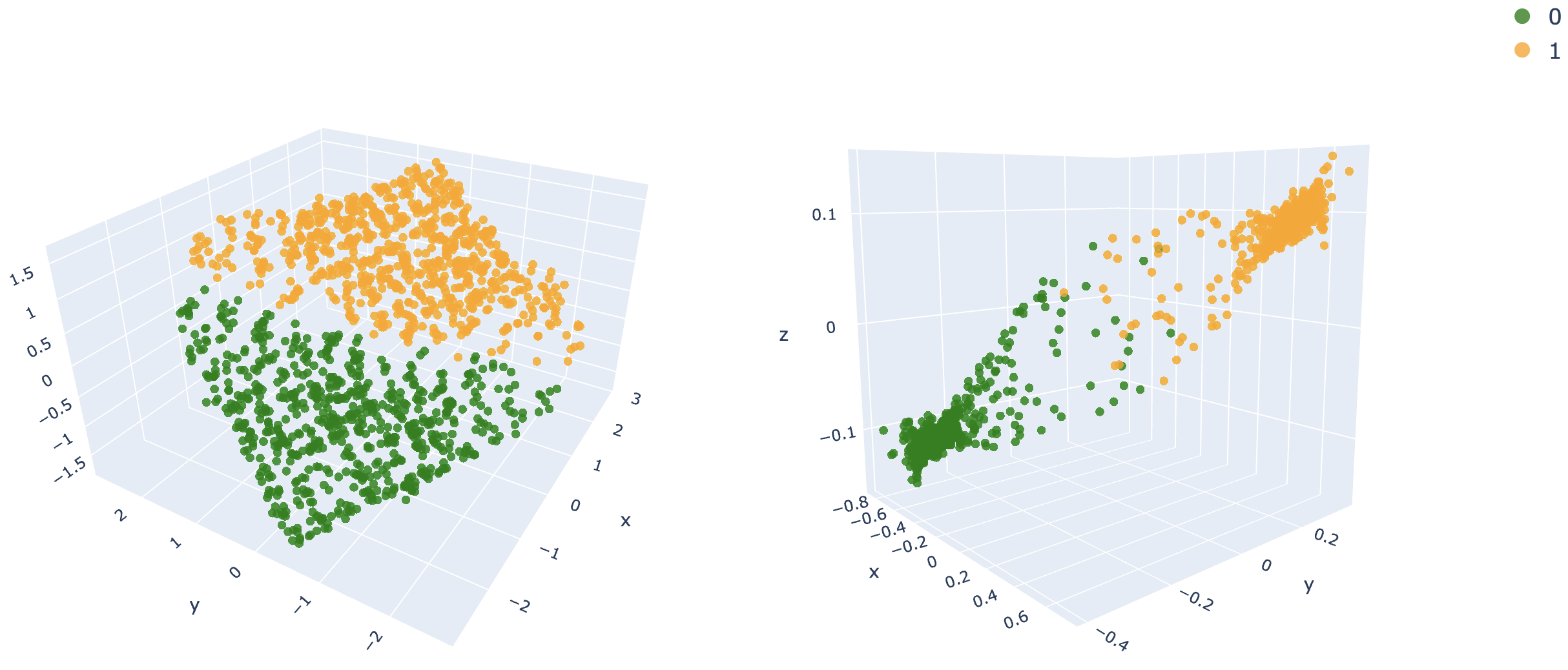}
\caption{
Comparison of the latent representations of the model E2M1, trained at high capacity (left) and low
capacity (right), in three dimensions.
}
\label{fig:E2M1_3D}
\end{center}
\end{figure}

%\newpage

\begin{figure}[h!]
\begin{center}
\includegraphics[width=\columnwidth]{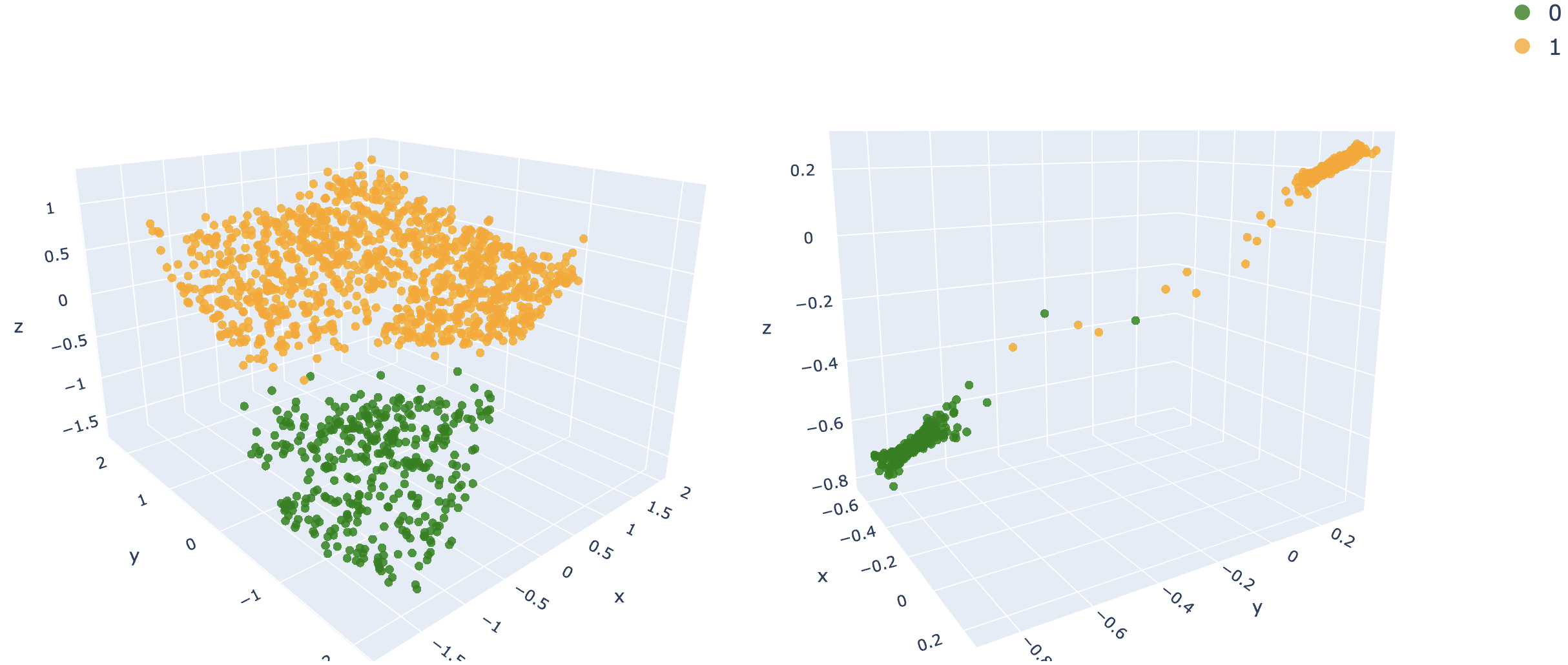}
\caption{
Comparison of the latent representations of the model E2M2, trained at high capacity (left) and low
capacity (right), in three dimensions.
}
\label{fig:E2M2_3D}
\end{center}
\end{figure}

%\newpage

\begin{figure}[h!]
\begin{center}
\includegraphics[width=\columnwidth]{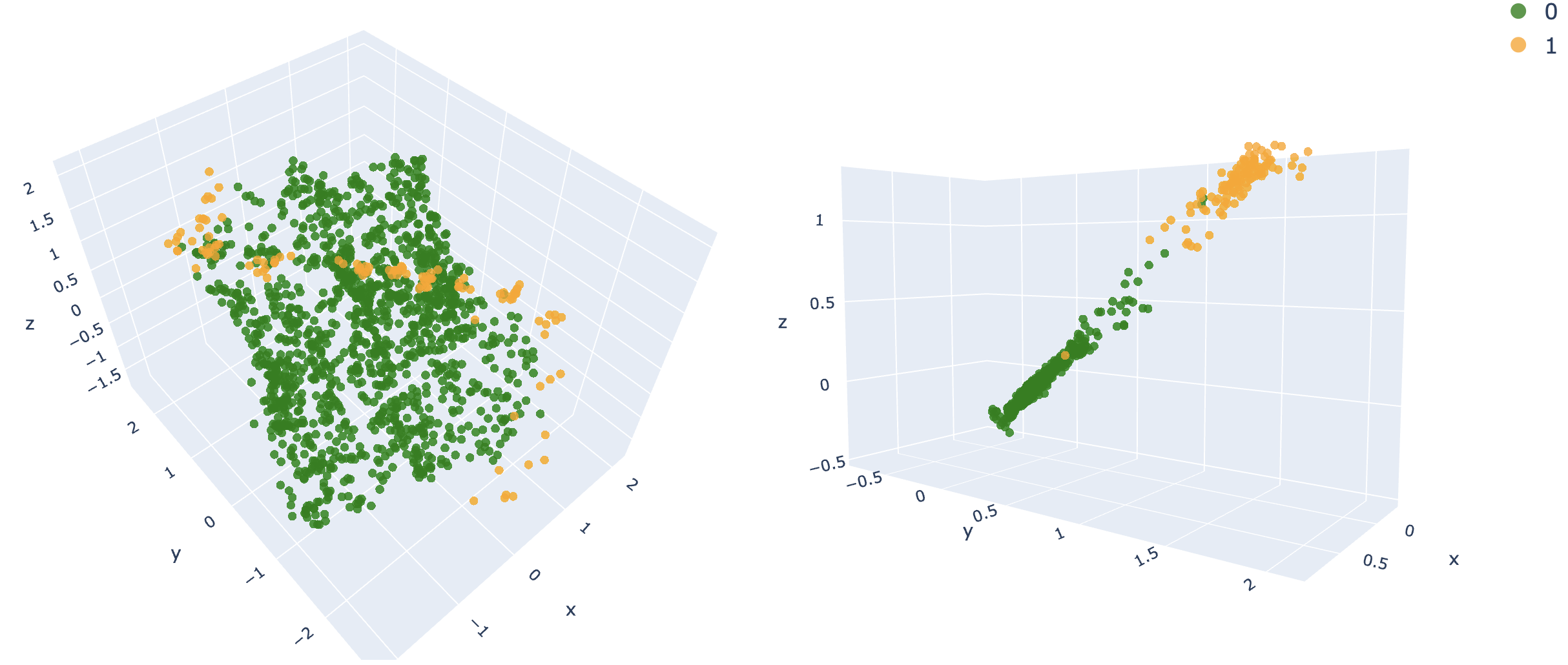}
\caption{
Comparison of the latent representations of the model E2M3, trained at high capacity (left) and low
capacity (right), in three dimensions.
}
\label{fig:E2M3_3D}
\end{center}
\end{figure}

%\newpage

\begin{figure}[h!]
\begin{center}
\includegraphics[width=\columnwidth]{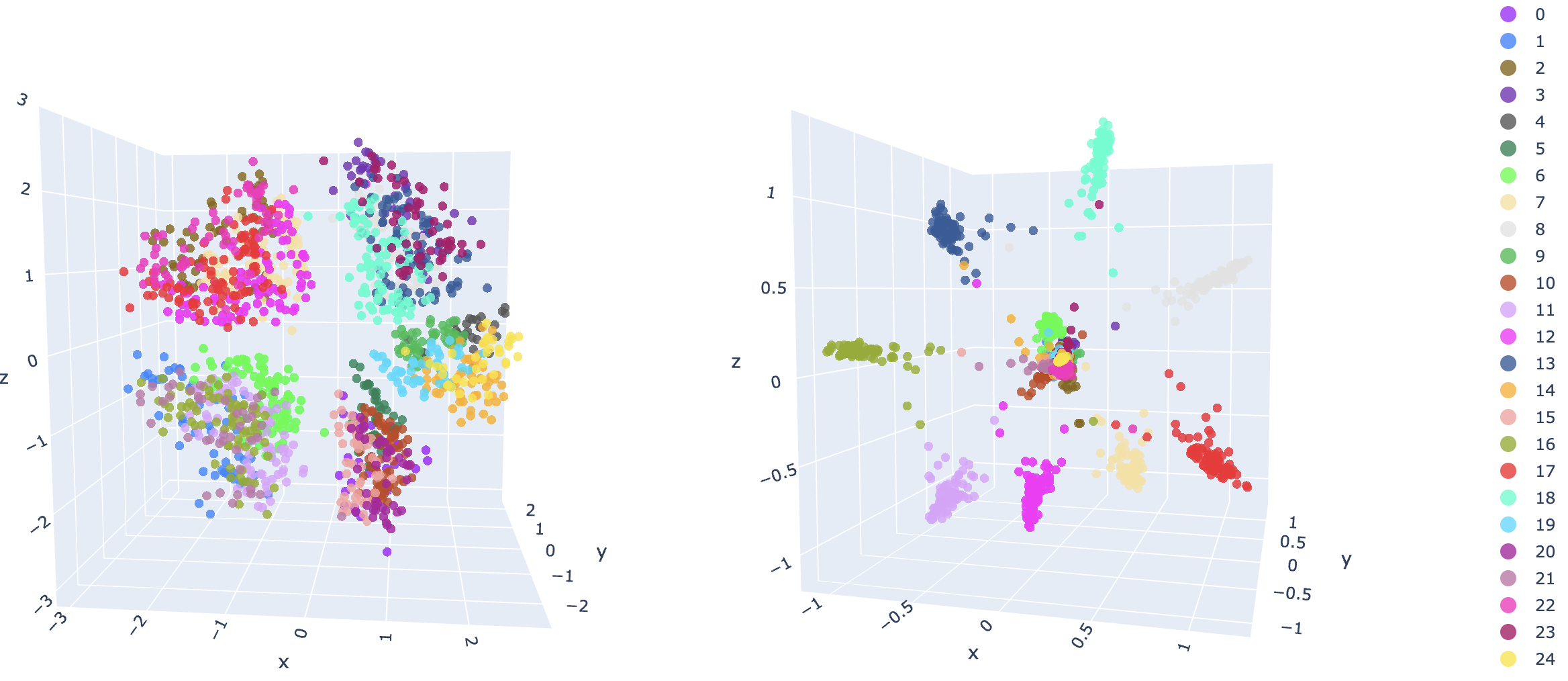}
\caption{
Comparison of the latent representations of the model E2M4, trained at high capacity (left) and low
capacity (right), in three dimensions.
}
\label{fig:E2M4_3D}
\end{center}
\end{figure}

%\newpage

\begin{figure}[h!]
\begin{center}
\includegraphics[width=\columnwidth]{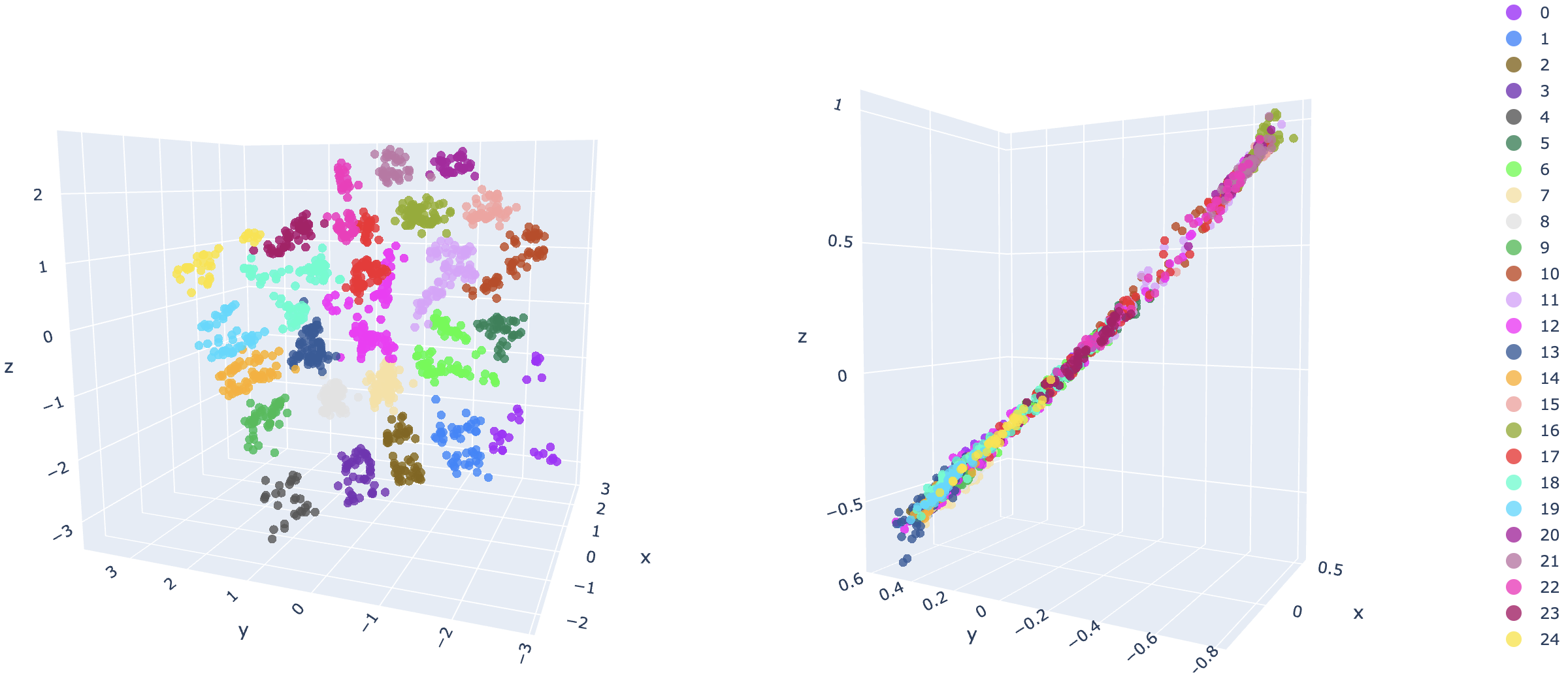}
\caption{
Comparison of the latent representations of the model E2M5, trained at high capacity (left) and low
capacity (right), in three dimensions.
}
\label{fig:E2M5_3D}
\end{center}
\end{figure}

\end{document}